\documentclass[superscriptaddress,onecolumn,amsmath,amssymb]{revtex4}

\usepackage{graphicx}
\usepackage{enumitem}

\newcommand{\sref}[1]{Sec.~\ref{#1}}
\newcommand{\aref}[1]{Appendix~\ref{#1}}
\newcommand{\eref}[1]{Eq.~(\ref{#1})}
\newcommand{\erefs}[1]{Eqs.~(\ref{#1})}

\newcommand{\rcite}[1]{Ref.~\cite{#1}}
\newcommand{\rcites}[1]{Refs.~\cite{#1}}

\usepackage{braket}
\newcommand{\ketbra}[2]{\mathinner{|{#1}\rangle \! \langle{#2}|}}
\newcommand{\dket}[1]{\mathinner{|{#1}\rangle \! \rangle}}
\newcommand{\dbra}[1]{\mathinner{\langle \! \langle {#1} |}}
\newcommand{\dbraket}[2]{\mathinner{\langle \! \langle {#1} | {#2} \rangle \! \rangle}}
\newcommand{\dketbra}[2]{\mathinner{|{#1} \rangle \! \rangle \! \langle \! \langle{#2}|}}

\usepackage{hyperref}
\hypersetup{colorlinks=true, linkcolor=blue, citecolor=blue, urlcolor=blue}

\begin{document}

\title{Exact diagonalization of a non-quadratic bosonic Liouvillian with two-body loss}

\author{Masaaki Tokieda}
\affiliation{Department of Chemistry, Graduate School of Science, Kyoto University, Kyoto 606-8502, Japan}

\begin{abstract}
We present the full diagonalization of a non-quadratic bosonic Liouvillian with a two-body loss term. The Liouvillian is shown to be exactly diagonalizable in terms of left and right confluent hypergeometric functions, whose distinction arises from the noncommutative nature of superoperators. The resulting spectral decomposition yields the general solution of the master equation, extending previous results. We further investigate the construction of a non-Gaussian open system model through the lens of nonlinear pseudomodes.
\end{abstract}

\maketitle

\section{Introduction}

The theory of open quantum systems serves not only to render quantum models more realistic by accounting for unavoidable dissipation effects but also to explore out-of-equilibrium phenomena in systems interacting with large reservoirs and to develop strategies for exploiting dissipation as a resource for quantum control.
Central to this framework is the master equation, which governs the evolution of the reduced system state \cite{Breuer02}.
Following several phenomenological studies of dissipation \cite{Bloch46,Senitzy60,Scully67}, the celebrated Markovian master equation was derived microscopically in the weak-coupling limit \cite{Davies}, and its most general mathematical structure was established through an axiomatic approach \cite{GKS,Lindblad}.
Although its validity as an approximation is seldom benchmarked against exact dynamics of underlying microscopic models, the Markovian master equation remains a widely used and effective description for systems weakly coupled to their bath.
Beyond its role as an approximation scheme, the Markovian master equation can also be employed to reproduce the reduced dynamics of microscopic Hamiltonians. This is achieved by introducing auxiliary degrees of freedom, so-called pseudomodes, that emulate the influence of the environment \cite{Imamoglu94, Garraway97, Tamascelli17, Smirne22,Tokieda25}.

Owing to its linear and autonomous structure, the general solution of a Markovian master equation can, in principle, be obtained by diagonalizing its generator, hereafter referred to as the Liouvillian.
While the spectrum or the kernel corresponding to steady states has been found even for complex systems \cite{Torres14,Bartolo16,Mamaev18,Nakagawa21}, achieving a full diagonalization, which is required to describe the dynamics, is challenging.
Among systems that admit full diagonalization, the most thoroughly studied are linear bosonic or fermionic systems, where the Liouvillian is quadratic in creation and annihilation operators, leading to linear equations of motion.
In such settings, full diagonalization was first achieved via an eigenvector ansatz \cite{Briegel93}.
It was later formulated systematically using Bogoliubov transformations of superoperators \cite{Prosen10,Guo17,Barthel22}, which have been extensively applied, for instance, in the study of topology in the presence of dissipation \cite{Lieu20,Flynn21,Nava23}. However, this method relies crucially on the quadratic structure and does not extend to nonlinear systems.
Few results are available in such cases, and existing approaches rely on identifying hidden structures that allow the Liouvillian to be reduced to a quadratic form in a suitable basis \cite{McDonald22}.

The main contribution of this work is to advance the diagonalization of non-quadratic Liouvillians.
We consider a single-mode bosonic system with a two-body loss term, which has been studied in various contexts, including dissipative stabilization of Schr\"odinger cat states \cite{Gilles94,Mazyar14,Zaki15}, the effects of strong dissipation in the Bose-Hubbard model such as fermionization \cite{Syassen08,Ripoll09} and the crossover to the Mott insulator phase \cite{Tomita17}, and the generation of nonclassical states for quantum metrology \cite{Munoz21,Karsa24}.
Previous work derived a limited set of eigenvectors using a truncated Fock basis \cite{Ripoll09}, whereas here we achieve the full diagonalization (\sref{diag}).
Our approach exploits symmetry considerations and a suitable change of basis, which render the Liouvillian block-diagonal and triangular within each block.
Analyzing the resulting eigenvectors, we find that the Liouvillian is diagonalized in terms of superoperator-valued confluent hypergeometric functions, with distinct left and right constructions arising from noncommutativity.
This structure goes beyond the linear regime, where diagonalization is achieved through Gaussian superoperators.
Using this result, we derive the general solution of the master equation and verify consistency with previous results (\sref{exact}).
Motivated by the goal of going beyond the Gaussian regime, we further investigate a non-Gaussian open system model by treating the non-quadratic Liouvillian as pseudomodes and propose a practical way to directly identify non-Gaussian features in the noise statistics (\sref{nonGauss}).
Conclusions are presented in \sref{conc}.

\section{Exact diagonalization}
\label{diag}

\subsection{Setups}

We consider a nonlinear oscillator mode under the influence of one- and two-body losses.
The annihilation and creation operators of the mode are denoted by $a$ and $a^\dagger \!$, respectively.
We also introduce the number operator $N = a^\dagger \! a$.
The density matrix $\rho(t)$ of the system follows $(d/dt) \rho(t) = \mathcal{L} \rho(t)$ with the Liouvillian $\mathcal{L}$ given by
\begin{equation}
    \mathcal{L} = - i \omega N^\times - \frac{iU}{2} (a^\dagger \! a^\dagger \! a a)^\times
    + \kappa_1 \mathcal{D}[a] + \kappa_2 \mathcal{D}[a^2],
    \label{eq:diag_L}
\end{equation}
where $\omega$ the eigenfrequency of the oscillator mode,
$U$ the strength of the nonlinear potential,
$\kappa_1$ and $\kappa_2$ the one- and two-body loss rates, respectively.
In the above equation, we have introduced the commutator superoperator $H^\times \bullet = H \bullet - \bullet H$ and the dissipator superoperator $\mathcal{D}[L] \bullet = L \bullet L^\dagger - (L^\dagger L \bullet + \bullet L^\dagger L)/2$, where $H$ and $L$ are operators.
The diagonalization of $\mathcal{L}$ for $\kappa_2 = 0$ was previously carried out in \rcite{McDonald22}. In the following sections, we extend this analysis by deriving the eigenvalues and eigenvectors for $\kappa_2 > 0$.
We note that the inclusion of the dephasing term $\mathcal{D}[N]$ does not alter the diagonalization procedure presented below.

To simplify notation, we introduce the vectorization of operators \cite{Havel03}.
Let $\dket{\rho}$ be a vector form of the operator $\rho$ such that $\dbra{\rho} \stackrel{\text{def}}{=} (\dket{\rho})^\dagger$ and $\dbraket{\rho_1}{\rho_2} = {\rm tr}(\rho_1^\dagger \rho_2)$ with ${\rm tr}$ the trace operation.
A superoperator becomes a matrix in the vectorized representation, which we denote by attaching the hat symbol $\hat{}$ as $\dket{\Phi \rho} = \hat{\Phi} \dket{\rho}$.

\subsection{Weak symmetry and spectrum}

The spectrum of $\mathcal{L}$ can readily be obtained using the weak symmetry consideration developed in \rcite{McDonald22}.
With $N = a^\dagger \! a$ the number operator, we first note the commutation relation $[\mathcal{L}, N^\times] = 0$.
This relation can be confirmed either by direct calculation or by the symmetry of $\mathcal{L}$.
The latter proceeds as follows.
From its definition, $\mathcal{L}$ is invariant under the phase rotation $a \to e^{i\theta} a \ (\theta \in \mathbb{R})$, which is given by $e^{i\theta} a = e^{-i \theta N} \, a \, e^{i \theta N} = e^{- i \theta N^\times} (a)$.
The invariance condition then reads $\mathcal{L}(  e^{i \theta N^\times} (\rho)) = e^{ i \theta N^\times} (\mathcal{L} (\rho))$ for any operator $\rho$, or equivalently $[\mathcal{L}, e^{i \theta N^\times}] = 0$.
Since this relation holds for any rotation angle $\theta$, we obtain $[\mathcal{L}, N^\times] = 0$.

The relation $[\mathcal{L}, N^\times] = 0$ indicates that $\mathcal{L}$ has a block diagonal structure in the eigenbasis of $N^\times$.
Fock states $\{ \ketbra{n_1}{n_2} \}_{n_1,n_2 \in \mathbb{Z}_{\geq 0}}$ are eigenvectors of $N^\times$.
To clarify the degeneracy, we introduce $\{ \phi_k^{(m)} \}_{m \in \mathbb{Z}, k \in \mathbb{Z}_{\geq 0}}$ as
\begin{equation*}
    \phi_k^{(m)} = \left\{
    \begin{array}{ll}
    \ketbra{k+m}{k} & (m \geq 0)\\
    \ketbra{k}{k-m} & (m < 0),
    \end{array}
    \right.
\end{equation*}
which satisfy $(\phi_k^{(m)})^\dagger = \phi_k^{(-m)}$,
the orthonormality condition $\dbraket{\phi_j^{(l)}}{\phi_k^{(m)}} = \delta_{l,m} \delta_{j,k}$ with the Kronecker delta $\delta$,
and the completeness relation $\hat{\mathcal{I}} = \sum_{m=-\infty}^{\infty} \sum_{k=0}^\infty \dketbra{\phi_k^{(m)}}{\phi_k^{(m)}}$ with the identity superoperator $\mathcal{I}$.
From the eigenvalue relation $\hat{N}^\times \dket{\phi_k^{(m)}} = m \dket{\phi_k^{(m)}}$, we conclude that,
in the basis $\{ \dket{\phi_k^{(m)}} \}_{m \in \mathbb{Z}, k \in \mathbb{Z}_{\geq 0}}$,
the Liouvillian $\hat{\mathcal{L}}$ has a block diagonal form, where each block is labeled by $m \in \mathbb{Z}$ and its elements are given by $\{ \dbra{\phi_j^{(m)}} \hat{\mathcal{L}} \dket{\phi_k^{(m)}} \}_{j,k \in \mathbb{Z}_{\geq 0}}$.

In fact, each block matrix is upper triangular.
To see this, we rewrite \eref{eq:diag_L} with the number operator $N$. Using $a^\dagger \! a^\dagger \! a a = N^2 - N$, we obtain
\begin{equation*}
    \mathcal{L} = - i \omega N^\times - \frac{\kappa_1}{2} N^\circ
    - \frac{iU}{2} (N^2 - N)^\times - \frac{\kappa_2}{2} (N^2 - N)^\circ
    + \kappa_1 \mathcal{A} + \kappa_2 \mathcal{A}^2,
\end{equation*}
where we have introduced the anti-commutator superoperator $H^\circ \bullet = H \bullet + \bullet H$ and $\mathcal{A} \bullet = a \bullet a^\dagger$.
The first four terms on the right hand side are diagonal in the basis $\{ \dket{\phi_k^{(m)}} \}$.
For the remaining terms, we note $\hat{\mathcal{A}}^j \dket{\phi_k^{(m)}} = j! \sqrt{\binom{k+|m|}{j} \binom{k}{j} } \dket{\phi_{k-j}^{(m)}} \ (k \geq j)$ and $= 0 \ (k < j)$, with $\binom{k}{j} = k!/(j!(k-j)!)$ the binomial coefficient.
Thus, the matrix element $\dbra{\phi_j^{(m)}} \hat{\mathcal{L}} \dket{\phi_k^{(m)}}$ is nonzero {\it iff} $j=k,k-1,$ or $k-2$, which indicates that each block is an upper triangular matrix.

To conclude, the basis $\{ \dket{\phi_k^{(m)}} \}$ puts $\hat{\mathcal{L}}$ into a block diagonal form, where each block matrix is upper triangular.
Thus, eigenvalues of $\mathcal{L}$ are given by the diagonal entries:
\begin{gather}
        \lambda_{k}^{(m)} = \dbra{\phi_k^{(m)}} \hat{\mathcal{L}} \dket{\phi_k^{(m)}} \nonumber \\
    \begin{gathered}
        = - i \Big( \omega - \frac{U}{2} \Big) m - (\kappa_1 + iUm) \Big( k + \frac{|m|}{2} \Big)
        - \kappa_2 \Big\{ (k+|m|)(k-1) + \frac{|m|(|m|+1)}{2} \Big\},
    \end{gathered}
    \label{eq:diag_eigval}
\end{gather}
for $m \in \mathbb{Z}$ and $k \in \mathbb{Z}_{\geq 0}$.
In addition to the relation $\lambda_k^{(-m)} = [\lambda_k^{(m)}]^*$, two remarks are in order.
\begin{enumerate}[label=(\alph*)]
  \item Dynamics in the asymptotic time regime ($t \to \infty$) are described by modes satisfying ${\rm Re}(\lambda_k^{(m)}) = 0$, where ${\rm Re}$ the real part.
  When $\kappa_1 > 0$, on the one hand, only one mode survives asymptotically, which is $(m,k) = (0,0)$ with $\lambda_0^{(0)} = 0$, resulting in a unique steady state.
  When $\kappa_1 = 0$, on the other hand, the condition is met by the four modes $(m,k) = (0,0)$, $(0,1)$, and $(\pm 1,0)$, the eigenvalues of which are $\lambda_0^{(0)} = \lambda_0^{(1)} = 0$ and $\lambda_0^{(\pm 1)} = \mp i \omega$, respectively.

  \item Note that
  \begin{equation*}
        \lambda_k^{(m)} - \lambda_q^{(m)} 
        = (q-k)(\kappa_1 + iUm + \kappa_2(|m|+q+k-1)).
  \end{equation*}
  When $\kappa_1 > 0$, we see that $\lambda_k^{(m)} \ne \lambda_q^{(m)} \ (k \ne q)$, which indicates that the eigenvalues in each block, $\{ \lambda_k^{(m)} \}_{k \in \mathbb{Z}_{\geq 0}}$, are all non-degenerate.
  On the other hand, when $\kappa_1 = 0$, as previously noted in the remark (a), we have $\lambda_0^{(0)} = \lambda_1^{(0)} = 0$, resulting in degeneracy in the block $m = 0$.
\end{enumerate}

\subsection{Eigenvectors}

Eigenvectors can be obtained by diagonalizing each block matrix.
This calculation can be facilitated by utilizing a similarity transformation with $e^{\mathcal{A}}$.
Notably, we show in \aref{app:sim.trans.} that $e^{\mathcal{A}} \mathcal{D}[a^2] e^{- \mathcal{A}} = \mathcal{A} (2 \mathcal{I} - N^\circ) - (1/2)(N^2-N)^\circ$ and that
\begin{equation}
    e^{\mathcal{A}} \mathcal{L} e^{- \mathcal{A}} = - i \omega N^\times  - \frac{\kappa_1}{2} N^\circ
    - \frac{iU}{2} (N^\circ - \mathcal{I}) N^\times  - \frac{\kappa_2}{2} (N^2 - N)^\circ
    - \mathcal{A} \{ \kappa_2 (N^\circ - 2 \mathcal{I}) + iU N^\times \}.
    \label{eq:diag_eALe-A}
\end{equation}
Consequently, $e^{\mathcal{A}} \mathcal{L} e^{- \mathcal{A}}$ is an upper triangular matrix, the elements of which are nonzero only on the diagonal and super-diagonal entries;
\begin{gather*}
    \dbra{\phi_j^{(m)}} e^{\mathcal{A}} \mathcal{L} e^{- \mathcal{A}} \dket{\phi_k^{(m)}} = \delta_{j,k-1} c_k^{(m)} + \delta_{j,k} \lambda_k^{(m)},
\end{gather*}
with $c_k^{(m)} = - \sqrt{k(k+|m|)} [ \kappa_2 \{ 2(k-1) + m \} + iUm ]$ and the eigenvalue $\lambda_k^{(m)}$.
The eigenvalue problem for this form of matrices can be solved analytically depending on the value of $\kappa_1/\kappa_2$ as follows;

\emph{Case (i)}, $\kappa_1/\kappa_2 \notin \mathbb{Z}_{\geq 0}$.
As noted in the remark (b) following \eref{eq:diag_eigval}, the eigenvalues in each block are all non-degenerate when $\kappa_1 > 0$.
The eigenvalue problem for such a matrix was solved, for instance, in \rcite{Torres19}.
As a result, we obtain $(\hat{\mathcal{L}} - \lambda_k^{(m)} \hat{\mathcal{I}} ) \dket{\rho_k^{(m)}} = 0$ and $\dbra{\bar{\rho}_k^{(m)}} (\hat{\mathcal{L}} - \lambda_k^{(m)} \hat{\mathcal{I}} )= 0$,
with $\dket{\rho_k^{(m)}}$ the right eigenvector given by
\begin{equation}
    e^{\hat{\mathcal{A}}} \dket{\rho_k^{(m)}} = \dket{\phi_k^{(m)}}
    + (1-\delta_{k,0}) \sum_{p = 0}^{k-1} \Big[ \prod_{q=p}^{k-1} \frac{c_{q+1}^{(m)}}{\lambda_k^{(m)} - \lambda_q^{(m)}} \Big] \dket{\phi_p^{(m)}},
    \label{eq:diag_Reigvec_0}
\end{equation}
and $\dbra{\bar{\rho}_k^{(m)}}$ the left eigenvector given by
\begin{equation}
    \dbra{\bar{\rho}_k^{(m)}} e^{- \hat{\mathcal{A}}} = \dbra{\phi_k^{(m)}}
    + \sum_{p = k+1}^{\infty} \Big[ \prod_{q=k+1}^{p} \frac{c_{q}^{(m)}}{\lambda_k^{(m)} - \lambda_q^{(m)}} \Big] \dbra{\phi_p^{(m)}},
    \label{eq:diag_Leigvec_0}
\end{equation}
for $m \in \mathbb{Z}$ and $k \in \mathbb{Z}_{\geq 0}$.

We can derive a more concise form of the eigenvectors, which proves beneficial in exploring their characteristics and in their applications.
The calculation proceeds in the following two steps (see \aref{app:eigvec.F.} for detailed derivation).
First, we evaluate the products inside the square brackets in \erefs{eq:diag_Reigvec_0} and (\ref{eq:diag_Leigvec_0}).
In this step, we utilize the relation $\prod_{q=l}^{l+n} (z+q) = (z+l)_{n+1}$ with $n,l \in \mathbb{Z}_{\geq 0}$, $z \in \mathbb{C}$, and $(z)_n$ the rising factorial defined by $(z)_0 = 1$ and $(z)_n = z (z+1) \cdots (z+n-1) \ (n > 0)$.
Second, we replace $\dket{\phi_p^{(m)}}$ in \eref{eq:diag_Reigvec_0} and $\dbra{\phi_p^{(m)}}$ in \eref{eq:diag_Leigvec_0} with $\dket{\phi_k^{(m)}}$ and $\dbra{\phi_k^{(m)}}$, respectively.
This can be achieved using the identities
$\hat{\mathcal{A}}^{k-p} \dket{\phi_k^{(m)}} = \sqrt{k!(k+|m|)!/(p!(p+|m|)!)} \dket{\phi_p^{(m)}}$ $(k \geq p)$ and
$\dbra{\phi_k^{(m)}} \hat{\mathcal{A}}^{p-k} = \sqrt{p!(p+|m|)!/(k!(k+|m|)!)} \dbra{\phi_p^{(m)}}$ $(p \geq k)$.
As a result, we obtain
\begin{equation}
    e^{\hat{\mathcal{A}}} \dket{\rho_k^{(m)}}
    =  \sum_{j = 0}^{k} \frac{(1 - x_k^{(m)})_j}{(2 - \{ 2x_k^{(m)} + \kappa_1/\kappa_2 \})_j} \frac{(2 \hat{\mathcal{A}})^{j}}{j!} \dket{\phi_k^{(m)}}
    \label{eq:diag_Reigvec_FiniteSum}
\end{equation}
and
\begin{equation}
    \dbra{\bar{\rho}_k^{(m)}} e^{- \hat{\mathcal{A}}} = 
    \dbra{\phi_k^{(m)}} {}_1 F_1(x_k^{(m)}; 2x_k^{(m)} + \frac{\kappa_1}{\kappa_2}; - 2 \hat{\mathcal{A}} ),
    \label{eq:diag_Leigvec_1}
\end{equation}
with $x_k^{(m)} = (2k+|m|)/2 + iUm/(2 \kappa_2)$ and ${}_1 F_1(\alpha;\beta;z) = \sum_{j=0}^\infty (\alpha)_j z^j / ((\beta)_j j!) \ (\alpha,\beta \in \mathbb{C}, \beta \notin \mathbb{Z}_{\leq 0})$ the confluent hypergeometric function.
The right eigenvectors \eref{eq:diag_Reigvec_FiniteSum} can be put in a more compact form when $\kappa_1 / \kappa_2 \notin \mathbb{Z}_{>0}$.
In this case, we have $2 - (2x_k^{(m)} + \kappa_1 / \kappa_2) \notin \mathbb{Z}_{\leq 0}$ for $m \in \mathbb{Z}$ and $k \in \mathbb{Z}_{\geq 0}$.
Then, using the relation $\hat{\mathcal{A}}^j \dket{\phi_k^{(m)}} = 0 \ (j > k)$, we can extend the upper limit of the sum from $k$ to $\infty$ and obtain
\begin{equation}
    e^{\hat{\mathcal{A}}} \dket{\rho_k^{(m)}}
    = {}_1 F_1 \Big( 1-x_k^{(m)};2 - \Big\{ 2x_k^{(m)} + \frac{\kappa_1}{\kappa_2} \Big\} ; 2 \hat{\mathcal{A}} \Big) \dket{\phi_k^{(m)}}.
    \label{eq:diag_Reigvec_1}
\end{equation}

Recall $\hat{N}^\times \dket{\phi_k^{(m)}} = m \dket{\phi_k^{(m)}}$, $\hat{N}^\circ \dket{\phi_k^{(m)}} = (2k+|m|) \dket{\phi_k^{(m)}}$, and that $\hat{N}^\times$ and $\hat{N}^\circ$ are Hermitian.
With $\mathcal{X} = N^\circ/2 + iU N^\times/(2 \kappa_2)$, thus, these relations lead to $\hat{\mathcal{X}} \dket{\phi_k^{(m)}} = x_k^{(m)} \dket{\phi_k^{(m)}}$ and $\dbra{\phi_k^{(m)}} \hat{\mathcal{X}} = x_k^{(m)} \dbra{\phi_k^{(m)}}$.
Eventually, we obtain
\begin{gather}
    e^{\hat{\mathcal{A}}} \dket{\rho_k^{(m)}}
    = \sum_{j = 0}^\infty \frac{ (2 \hat{\mathcal{A}})^j}{j!} \frac{(\hat{\mathcal{I}}-\hat{\mathcal{X}})_j}{ \Big( 2 \hat{\mathcal{I}} - \Big\{ 2 \hat{\mathcal{X}} + \frac{\kappa_1}{\kappa_2} \hat{\mathcal{I}} \Big\} \Big)_j } \dket{\phi_k^{(m)}} \nonumber \\
    \begin{gathered}
        \stackrel{\text{def}}{=} {}_1 F_1^R \Big( \hat{\mathcal{I}}-\hat{\mathcal{X}}; 2 \hat{\mathcal{I}} - \Big\{ 2 \hat{\mathcal{X}} + \frac{\kappa_1}{\kappa_2} \hat{\mathcal{I}} \Big\} ; 2 \hat{\mathcal{A}} \Big) \dket{\phi_k^{(m)}},
    \end{gathered}
    \label{eq:diag_Reigvec}
\end{gather}
and
\begin{gather}
    \dbra{\bar{\rho}_k^{(m)}} e^{- \hat{\mathcal{A}}} 
        = \dbra{\phi_k^{(m)}} \sum_{j = 0}^\infty \frac{(\hat{\mathcal{X}})_j}{ \Big( 2 \hat{\mathcal{X}} + \frac{\kappa_1}{\kappa_2} \hat{\mathcal{I}} \Big)_j } \frac{ ( - 2 \hat{\mathcal{A}})^j}{j!} \nonumber \\
    \begin{gathered}
        \stackrel{\text{def}}{=} \dbra{\phi_k^{(m)}} {}_1 F_1^L \Big( \hat{\mathcal{X}};2 \hat{\mathcal{X}} + \frac{\kappa_1}{\kappa_2} \hat{\mathcal{I}} ; - 2 \hat{\mathcal{A}} \Big),
    \end{gathered}
    \label{eq:diag_Leigvec}
\end{gather}
where the rising factorial has been extended to include operator arguments as $(\hat{\mathcal{X}})_n = \hat{\mathcal{X}} (\hat{\mathcal{X}} + \hat{\mathcal{I}}) \cdots (\hat{\mathcal{X}} + (n-1)\hat{\mathcal{I}})$.
It should be noted that the placement of the rising factorials differs between the definition of ${}_1 F_1^R$ and ${}_1 F_1^L$:
While the rising factorials appear on the right side in the definition of ${}_1 F_1^R$, they appear on the left side in ${}_1 F_1^L$.
The non-commutative character of $\mathcal{A}$ and $\mathcal{X}$ results in a distinction between the two.

Now, let $\mathcal{F} = {}_1 F_1^L ( \mathcal{X};2 \mathcal{X} + (\kappa_1/\kappa_2) \mathcal{I} ; - 2 \mathcal{A} )$.
Notably, we can prove that the inverse of $\mathcal{F}$, $\mathcal{F}^{-1}$, is given by (see \aref{app:F-1} for the proof)
\begin{equation}
    \mathcal{F}^{-1} = {}_1 F_1^R \Big( \mathcal{I} - \mathcal{X}; 2 \mathcal{I} - \Big\{ 2 \mathcal{X} + \frac{\kappa_1}{\kappa_2} \mathcal{I} \Big\} ; 2 \mathcal{A} \Big).
    \label{eq:diag_F-1}
\end{equation}
In other words,
\begin{equation}
    \dket{\rho_k^{(m)}} = ( \hat{\mathcal{F}} e^{ \hat{\mathcal{A}}} )^{-1} \dket{\phi_k^{(m)}}, \ \  \dbra{\bar{\rho}_k^{(m)}} =  \dbra{\phi_k^{(m)}} \hat{\mathcal{F}} e^{ \hat{\mathcal{A}}}.
    \label{eq:diag_rho}
\end{equation}
On the one hand, the right inverse relation $\mathcal{F} \mathcal{F}^{-1} = \mathcal{I}$ leads to the orthonormality condition $\dbraket{\bar{\rho}_j^{(l)}}{\rho_k^{(m)}} = \delta_{l,m} \delta_{j,k} $.
On the other hand, the left inverse relation $\mathcal{F}^{-1} \mathcal{F} = \mathcal{I}$ yields the completeness relation $\hat{\mathcal{I}} = \sum_{m=-\infty}^\infty \sum_{k=0}^\infty \dketbra{\rho_k^{(m)}}{\bar{\rho}_k^{(m)}}$,
which ensures that $\{ \dket{\rho_k^{(m)}}, \dbra{\bar{\rho}_k^{(m)}} \}_{m \in \mathbb{Z}, k \in \mathbb{Z}_{\geq 0}}$ is a complete set of eigenvectors.
Note that \erefs{eq:diag_rho} further yield
\begin{equation*}
    (\mathcal{F} e^{\mathcal{A}}) \, \mathcal{L} \, (\mathcal{F} e^{\mathcal{A}})^{-1} = - i \omega N^\times - \frac{\kappa_1}{2} N^\circ
    - \frac{iU}{2} (N^2 - N)^\times - \frac{\kappa_2}{2} (N^2 - N)^\circ
\end{equation*}

From $(x_k^{(m)})^* = x_k^{(-m)}$ and $[\mathcal{A} (\rho)]^\dagger = \mathcal{A} (\rho^\dagger)$, we can show $(\rho_k^{(m)})^\dagger = \rho_k^{(-m)}$ and $(\bar{\rho}_k^{(m)})^\dagger = \bar{\rho}_k^{(-m)}$.
We provide the explicit forms of several eigenvectors in \aref{app:exp.eigvec.}.

In the limit $\kappa_2 \to 0$, we obtain $\mathcal{F} e^{\mathcal{A}} = \exp \left(\mathcal{A}/(1 + i (U/\kappa_1) N^\times)\right)$.
The Liouvillian can then be diagonalized blockwise in $m$ using the Gaussian superoperator $\exp \left(\mathcal{A}/(1 + i m U/\kappa_1)\right)$.
This also follows from \eref{eq:diag_eALe-A}, since the problem reduces to a quadratic form within each block $m$, where $N^\times$ is replaced by the scalar $m$, as pointed out in \rcite{McDonald22}.
The system thus remains effectively linear in this limit, whereas the confluent hypergeometric function derived here captures genuinely nonlinear behavior beyond this regime.

\emph{Case (ii)}, $\kappa_1/\kappa_2 \in \mathbb{Z}_{> 0}$.
In this case, $2 - (2x_k^{(m)} + \kappa_1 / \kappa_2)$ can be a non-positive integer.
Accordingly, the right eigenvectors are given by the finite sum in \eref{eq:diag_Reigvec_FiniteSum}.

\emph{Case (iii)}, $\kappa_1 = 0$.
In this case, as noted in the remark (b), the block $m = 0$ contains degeneracy as $\lambda_0^{(0)} = \lambda_1^{(0)} = 0$.
Consequently, we cannot apply the formulas \eref{eq:diag_Reigvec_0} for $k=0$ and $1$ and \eref{eq:diag_Leigvec_0} for $k=0$ straightforwardly.
Regarding this matter, it is worth noting that $c_1^{(0)} = 0$. This implies that the block matrix with $m=0$ is in a block diagonal form as
\begin{equation*}
  \left[
  \begin{array}{c|cccc}
    0 & 0 & 0 & 0 & \cdots \\ \hline
    0 & 0 & c_2^{(0)} & 0 & \cdots \\
    0 & 0 & \lambda_2^{(0)} & c_3^{(0)} & \cdots \\
    0 & 0 & 0 & \lambda_3^{(0)} & \cdots \\
    & & \vdots \\
  \end{array}
  \right].
\end{equation*}
As there is no degeneracy in the lower block matrix, the eigenvectors for this block can be expressed in a similar manner to \erefs{eq:diag_Reigvec_0} and (\ref{eq:diag_Leigvec_0}).
Following the calculation procedures presented above, we obtain
\begin{equation*}
  e^{\hat{\mathcal{A}}} \dket{\rho_k^{(m)}}
  =  \sum_{j = 0}^{k} \frac{(1 - x_k^{(m)})_j}{(2 \{ 1 - x_k^{(m)} \} )_j} \frac{(2 \hat{\mathcal{A}})^{j}}{j!} \dket{\phi_k^{(m)}},
\end{equation*}
and
\begin{equation*}
  \dbra{\bar{\rho}_k^{(m)}} e^{- \hat{\mathcal{A}}}
  = \dbra{\phi_k^{(m)}} {}_1 F_1(x_k^{(m)}; 2x_k^{(m)}; - 2 \hat{\mathcal{A}} ),
\end{equation*}
for all the modes except $(m,k) = (0,0)$ and $(0,1)$. The modes corresponding to $(m,k) = (0,0)$ and $(0,1)$ have the eigenvalue 0 and are 2-fold degenerate.
The associated right eigenspace is spanned by the vectors 
\begin{equation*} 
    e^{-\hat{\mathcal{A}}} \dket{\phi_0^{(0)}} = \dket{\phi_0^{(0)}} \ \ {\rm and} \ \
    e^{-\hat{\mathcal{A}}} \dket{\phi_1^{(0)}} = \dket{\phi_1^{(0)}} - \dket{\phi_0^{(0)}},
\end{equation*}
and the left eigenspace by (see \erefs{eq:exp.eigvec._L_0^0} and (\ref{eq:exp.eigvec._L_1^0}) for the derivations)
\begin{equation*}
    \dbra{\phi_0^{(0)}} e^{\hat{\mathcal{A}}} = \dbra{I} \ \ {\rm and} \ \ 
    \dbra{\phi_1^{(0)}} e^{\hat{\mathcal{A}}} {}_1 F_1(1;2;- 2 \hat{\mathcal{A}} ) = \dbra{ I - \Pi_+ },
\end{equation*}
with $\Pi_+ = \sum_{n = 0}^\infty \ketbra{2n}{2n}$ the parity operator.

As noted in the remark (a), when $\kappa_1 = 0$, only the modes $(m,k) = (0,0)$, $(0,1)$, and $(\pm 1,0)$ survive in the asymptotic time regime.
For the modes $(m,k) = (0,0)$ and $(0,1)$, a possible choice of the eigenvectors that respects the orthonormality is
\begin{gather*}
    \rho_0^{(0)} = \ketbra{0}{0}, \ \ \bar{\rho}_0^{(0)} = \Pi_+, \\
    \rho_1^{(0)} = \ketbra{1}{1}, \ \ \bar{\rho}_1^{(0)} = I - \Pi_+.
\end{gather*}
Assuming $U = 0$, the eigenvectors of the modes $(m,k) = (\pm 1,0)$ are given by (see \erefs{eq:exp.eigvec._R_0^m} and (\ref{eq:exp.eigvec._L_0^m}) for the derivations)
\begin{gather*}
    \rho_0^{(1)} = \ketbra{1}{0}, \ \ \bar{\rho}_0^{(1)} = a^\dagger \Pi_+ \frac{ (N-1)!! }{(N)!!}, \\
    \rho_0^{(-1)} = (\rho_0^{(1)})^\dagger, \ \ \bar{\rho}_0^{(-1)} = (\bar{\rho}_0^{(1)})^\dagger,
\end{gather*}
with $k!! = k \times (k-2)!! \ (k \in \mathbb{Z}_{\geq -1})$ the double factorial ($0!! = (-1)!! \stackrel{\text{def}}{=} 1$).
These are consistent with the results obtained in Appendix A.1 of \rcite{Mazyar14} when the two-photon driving is neglected.

\section{Exact solution of the master equation}
\label{exact}

Using the results of the previous section, here we derive the general solution of the Markovian master equation $\dot{\rho}(t)=\mathcal{L}\rho(t)$ and verify its consistency with earlier results.
Given the full set of eigenvalues and eigenvectors, we can perform the spectral decomposition of the propagator $e^{\hat{\mathcal{L}} t}$ as
\begin{equation}
    e^{\hat{\mathcal{L}} t} = \sum_{m = - \infty}^\infty \sum_{k = 0}^\infty e^{\lambda_k^{(m)} t} \dketbra{\rho_k^{(m)}}{\bar{\rho}_k^{(m)}}.
    \label{eq:exact_spec.decomp.}
\end{equation}
We can then express the formal solution of the master equation $\rho(t) = e^{\mathcal{L} t} \rho(0)$ as
\begin{equation}
    \rho(t) = \sum_{m = - \infty}^\infty \sum_{k = 0}^\infty b_k^{(m)} e^{\lambda_k^{(m)} t} \rho_k^{(m)},
    \label{eq:exact_formal.solution}
\end{equation}
where the expansion coefficients are calculated from the left eigenvectors $b_k^{(m)} = \dbraket{\bar{\rho}_k^{(m)}}{\rho(0)}$.
This expression is convenient if the initial state is a coherent state, $\rho(0) = \ketbra{\alpha}{\alpha}$ with $\ket{\alpha} = e^{\alpha a^\dagger - \alpha^* a} \ket{0}$ because $(\mathcal{A}_{-} - |\alpha|^2 \mathcal{I}) \ketbra{\alpha}{\alpha} = 0$ and we can evaluate $b_k^{(m)}$ using \eref{eq:diag_Leigvec} as
\begin{gather*}
    b_k^{(m)} = \frac{|\alpha|^{|m|+2k} e^{im\varphi} }{\sqrt{k! (|m|+k)!}} {}_1 F_1 ( x_k^{(m)}; 2x_k^{(m)} + \kappa_1/\kappa_2 ; - 2 |\alpha|^2),
\end{gather*}
where we have assumed $\alpha = |\alpha| e^{i \varphi}$.

The solution can also be expressed in the Fock basis as 
\begin{equation*}
    \dbraket{\phi_k^{(m)}}{\rho(t)} = \sum_{\mu = - \infty}^\infty \sum_{q = 0}^\infty \dbra{\phi_k^{(m)}} e^{\hat{\mathcal{L}} t} \dket{\phi_q^{(\mu)}} \dbraket{\phi_q^{(\mu)}}{\rho(0)}.
\end{equation*}
The matrix elements $\dbra{\phi_k^{(m)}} e^{\hat{\mathcal{L}} t} \dket{\phi_q^{(\mu)}}$ can be evaluated by inserting the spectral decomposition \eref{eq:exact_spec.decomp.}. 
This is implemented in \aref{app:time.evolv.} (see \eref{eq:time.evolv._prop.me}). As a result, we obtain
\begin{equation}
    \dbraket{\phi_k^{(m)}}{\rho(t)} = \sum_{r = 0}^\infty \sqrt{ \binom{|m|+k+r}{|m|+k}\binom{k+r}{k}}  G_{r,k}^{(m)}(t) \dbraket{\phi_{k+r}^{(m)}}{\rho(0)},
    \label{eq:exact_time.evolv.}
\end{equation}
with
\begin{gather*}
    G_{r,k}^{(m)}(t) = \sum_{j = 0}^r (-1)^j \binom{r}{j} e^{\lambda_{k+j}^{(m)} t} \\
    \times {}_2 F_1 \Big( -j, 1 - x_{k+j}^{(m)}; 2 - 2 x_{k+j}^{(m)} - \frac{\kappa_1}{\kappa_2}; 2 \Big)
    {}_2 F_1 \Big( -(r-j), x_{k+j}^{(m)}; 2 x_{k+j}^{(m)} + \frac{\kappa_1}{\kappa_2}; 2 \Big).
\end{gather*}
For $\omega = U = \kappa_1 = 0$ ($\mathcal{L} = \kappa_2 \mathcal{D}[a^2]$), the exact solution of the master equation was found in \rcites{Simaan75,Simaan78} by solving the differential equation using the method of a generating function.
In \aref{app:time.evolv.}, we show that \eref{eq:exact_time.evolv.} is consistent with their result.

From the solution $\rho(t) = e^{\mathcal{L} t} \rho(0)$, the expectation value of an operator $O$ is calculated by ${\rm tr} (\rho(t) O)$.
Using the Hermitian property of $\rho(t)$, this can be expressed as
\begin{gather*}
    {\rm tr} (\rho(t) O) = \dbraket{\rho(t)}{O} = \dbraket{e^{\mathcal{L} t} \rho(0)}{O} \\
    = \dbra{\rho(0)} e^{\hat{\mathcal{L}}^\dagger t} \dket{O}.
\end{gather*}
The second line indicates that $\dket{O^{H} (t)} \stackrel{\text{def}}{=} e^{\hat{\mathcal{L}}^\dagger t} \dket{O}$ is the Heisenberg representation of the operator $O$.
By taking the Hermitian conjugate of \eref{eq:exact_spec.decomp.}, the spectral decomposition of $e^{\hat{\mathcal{L}}^\dagger t}$ reads as 
\begin{equation*}
    e^{\hat{\mathcal{L}}^\dagger t} = \sum_{m = - \infty}^\infty \sum_{k = 0}^\infty e^{\lambda_k^{(-m)} t} \dketbra{\bar{\rho}_k^{(m)}}{\rho_k^{(m)}}.
\end{equation*}
This yields the formal expression of $O^{H}(t)$;
\begin{equation*}
    O^{H}(t) = \sum_{m = - \infty}^\infty \sum_{k = 0}^\infty \bar{b}_k^{(m)} e^{\lambda_k^{(-m)} t} \bar{\rho}_k^{(m)},
\end{equation*}
where, this time, the expansion coefficients are calculated from the right eigenvectors $\bar{b}_k^{(m)} = \dbraket{\rho_k^{(m)}}{O}$.

The expression in the Fock basis can similarly be obtained as (see \aref{app:time.evolv.} for the derivation)
\begin{equation*}
    \dbraket{\phi_k^{(m)}}{O^H (t)} = \sum_{q = 0}^k \sqrt{ \binom{|m|+k}{|m|+q}\binom{k}{q}} G_{k-q,q}^{(-m)}(t) \dbraket{\phi_{q}^{(m)}}{O}.
\end{equation*}
For the annihilation operator $O = a$, it follows that $a = \sum_{k = 0}^\infty \sqrt{k+1} \, \ketbra{k}{k+1}$ and $\dbraket{\phi_{q}^{(m)}}{a} = \delta_{m,-1} \sqrt{q+1}$. This leads to 
\begin{equation*}
    a^H (t) = \sum_{k = 0}^\infty \left\{ \sum_{q=0}^k \binom{k}{q} G_{k-q,q}^{(1)}(t) \right\} \sqrt{k+1} \, \ketbra{k}{k+1}.
\end{equation*}
Hence, the matrix form of $a$ remains invariant under the time evolution.
This is a consequence of the weak symmetry. 
The time evolution of the element in the $k$-th row can be obtained by multiplying the factor $\sum_{q=0}^k \binom{k}{q} G_{k-q,q}^{(1)}(t)$.

\section{Characterization of Non-Gaussian open system dyanmics}
\label{nonGauss}

Here we investigate non-Gaussian open system dynamics by treating a nonlinear bosonic mode as a pseudomode.
Consider a system coupled to an environment via the interaction Hamiltonian $V_{SE} = V_S \otimes V_E$, where $V_S$ ($V_E$) is a Hermitian operator acting on the system (environment).
The total density operator $\rho_{SE}(t)$ evolves as $(d/dt) \rho_{SE}(t) = [\mathcal{L}_S + \mathcal{L}_E - (i/\hbar) \sum_{p = \pm} p V_{SE}^p] \rho_{SE}(t)$, 
where $\mathcal{L}_S$ and $\mathcal{L}_E$ generate internal dynamics and $V_{SE}^\pm = (V_{SE}^\circ \pm V_{SE}^\times)/2$.
For an initially factorized state $\rho_{SE}(0) = \rho_S(0) \otimes \rho_E$, the reduced system dynamics is fully characterized by the set of multi-time correlation functions of the environment
\begin{equation}
    C^{(n)}_{p_1,\cdots,p_n} (t_1, \cdots, t_n) = {\rm tr}_E \left[ \widetilde{V_E^{p_1}}(t_1) \cdots \widetilde{V_E^{p_n}}(t_n) \rho_E \right],
    \label{eq:nonGauss_corr}
\end{equation}
where $p_k=\pm$ and $t_1 \ge \cdots \ge t_n$, ${\rm tr}_E$ denotes the trace over the environment, and $\widetilde{V_E^p}(t) = e^{-\mathcal{L}_E t} \, V_E^p \, e^{\mathcal{L}_E t}$.
For a linear bosonic system initialized in a Gaussian state, all higher-order correlation functions can be expressed in terms of the first two orders ($n=1,2$) owing to Wick's theorem.
This structure implies Gaussian statistics of the underlying noise \cite{Weiss08,Diosi14}.
Most analyses of open system dynamics have been confined to this Gaussian regime, despite the growing theoretical and experimental interest in non-Gaussian effects \cite{Wen13,Sung19,Ferioli24}.
To go beyond Gaussianity, one may introduce either a non-Gaussian initial state $\rho_E$ or
nonlinear environmental dynamics through $\widetilde{V_E^p}(t)$. 
Here, we examine the latter approach, taking $\mathcal{L}_E$ as in \eref{eq:diag_L} and $V_E = a + a^\dagger$.
Employing a Markovian master equation for $\mathcal{L}_E$ provides a practical alternative to Hamiltonian environment models, which require many degrees of freedom to capture irreversibility. 

A non-Gaussian open-system model based on a Markovian master equation was studied using two-level systems in \rcite{Funo24}. The authors derived analytic expressions for multi-time correlation functions akin to \eref{eq:nonGauss_corr} and revealed Poissonian noise statistics.
For the nonlinear bosonic system, analytic expressions can likewise, in principle, be obtained from the spectral decomposition \eref{eq:exact_spec.decomp.}.
However, these expressions are typically too cumbersome to provide clear physical insight.
This motivates the need for a practical and systematic framework to characterize higher-order correlation functions, in particular their non-Gaussian features.

To this end, it might be convenient to introduce the generating functional
\begin{equation*}
    Z[J_+, J_-] = {\rm tr}_E \left[ \mathcal{T} \exp\left( i \int_0^t d\tau \sum_{p = \pm} J_p(\tau) \widetilde{V_E^p}(\tau) \right)  \rho_E \right],
\end{equation*}
where $\mathcal{T}$ denotes chronological time ordering.
Since its functional derivatives generate the correlation functions \cite{Felix17}, the functional $Z[J_+, J_-]$ uniquely characterizes the influence of the environment.
Introducing the environment operator $\xi_E(t)$ defined by
\begin{equation*}
    \frac{d}{dt} \xi_E(t) = \left[ \mathcal{L}_E + i \sum_{p = \pm} J_p(t) V_E^p \right] \xi_E(t), \ \ \ \xi_E(0) = \rho_E,
\end{equation*}
the generating functional can be evaluated as $Z[J_+, J_-] = {\rm tr}_E [\xi_E(t)]$.

\begin{figure}[t]
\includegraphics[keepaspectratio, scale=0.7]{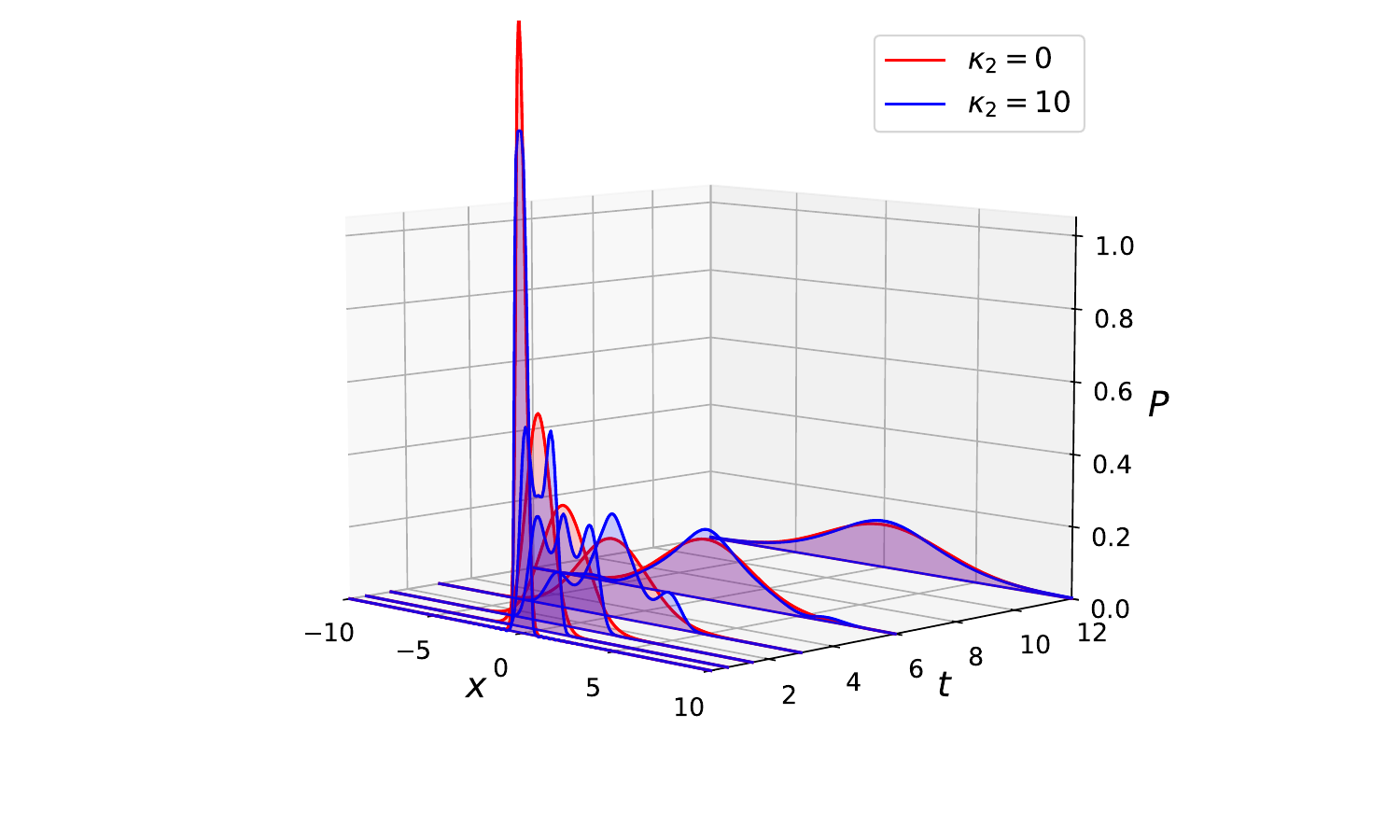}
\caption{
Comparison of the probability density $P(x)$ for the linear ($\kappa_2 = 0$, red) and nonlinear ($\kappa_2 = 10$, blue) cases.
Here $\omega = 1$ and $U = 0$ in the unit $\kappa_1 = 1$, and the system is initialized in the vacuum state.
}
\label{fig1}
\end{figure}

The full functional $Z[J_+, J_-]$ is quite involved.
To illustrate non-Gaussian behavior, we restrict to constant sources $J_+(t)=J_-(t)=J/2$.
In this case, the generating functional $Z(J)$ reduces to a characteristic function, whose associated probability density
\begin{equation*}
    P(x) = \int_{-\infty}^{\infty} \frac{dJ}{2\pi} Z(J) e^{-iJx},
\end{equation*}
generates the time-integrated noise correlations
\begin{equation*}
    \int_{-\infty}^\infty dx P(x) x^n
    = \frac{n!}{2^n} \int_0^t dt_1 \dots \int_0^{t_{n-1}} dt_n {\rm tr_B} \left[ \widetilde{V_E^{\circ}}(t_1) \cdots \widetilde{V_E^{\circ}}(t_n) \rho_E \right].
\end{equation*}
While this choice sacrifices time resolution, non-Gaussianity can be inferred from deviations of $P(x)$ from a Gaussian form.
The numerical evolution of the probability density $P(x)$ from the initial vacuum state is illustrated in Fig.~\ref{fig1}. While the linear system maintains a Gaussian profile at all times (red curve), the nonlinear dynamics (blue curve) transform the initial singular distribution into a multimodal structure, reflecting the strong non-Gaussianity inherent in the pseudomode. Despite this pronounced transient behavior, the distribution restores its Gaussian form in the long-time limit. This convergence is driven by the clustering property of the higher-order correlation functions, which ensures the extensivity of the cumulants and validates the application of the central limit theorem \cite{Touchette09}. We confirm through explicit calculation that the fourth-order cumulants decay as the time arguments become widely separated.

\section{Conclusion}
\label{conc}

In this work, we achieved the full diagonalization of a non-quadratic bosonic Liouvillian describing a single-mode system under the influence of two-body loss. By exploiting weak symmetry and an appropriate basis transformation, the Liouvillian was reduced to a block-diagonal and triangular form, allowing us to obtain its complete spectrum analytically. The corresponding left and right eigenvectors were constructed in closed form in terms of superoperator-valued confluent hypergeometric functions, reflecting the intrinsically noncommutative structure of the problem. This formulation extends the well-known Gaussian framework and captures genuinely nonlinear effects beyond it. Based on the spectral decomposition, we derived the general solution of the Markovian master equation and expressed the dynamics explicitly for arbitrary initial states. The solution was further validated by demonstrating consistency with previously known exact results in limiting cases. 
Beyond the diagonalization itself, we explored the construction of a non-Gaussian open system model using a nonlinear pseudomode, in line with the central motivation of this work to go beyond the Gaussian regime. We proposed the use of a generating functional as a practical tool to identify non-Gaussian features in the noise statistics. By considering constant source functions, which yield the statistics of time-integrated correlation functions, we demonstrated the emergence of pronounced non-Gaussian features in the transient dynamics, followed by a restoration of Gaussian behavior due to the clustering of higher-order correlations.
This work opens several directions for future research.
While our derivation of the confluent hypergeometric functions relied on direct algebraic manipulation, developing a more systematic framework, analogous to Bogoliubov transformation for linear systems, would significantly broaden the class of tractable non-quadratic Liouvillians.
Furthermore, while we restricted our pseudomode analysis to constant source functions, a more tailored choice of sources could enable selective probing of higher-order correlation functions \cite{Ono25} and provide a more detailed characterization of non-Gaussian noise statistics.

\begin{acknowledgments}
    I thank Pierre Rouchon for his valuable input. This work was supported by the Engineering for Quantum Information Processors (EQIP) Inria challenge project and JSPS KAKENHI Grant Number JP23KJ1157.
\end{acknowledgments}

\appendix

\section{Similarity transformations with $e^{\mathcal{A}}$}
\label{app:sim.trans.}

In this appendix, we consider similarity transformations with $e^{\mathcal{A}}$ to derive \eref{eq:diag_eALe-A}.
Using $[\mathcal{A}, N^\times] = 0$, $[\mathcal{A}, N^\circ] = 2 \mathcal{A}$, and
$e^{\mathcal{A}} \bullet e^{- \mathcal{A}} = \bullet + [\mathcal{A}, \bullet] + [\mathcal{A} , [\mathcal{A}, \bullet] ]/2! + [ \mathcal{A}, [\mathcal{A} , [\mathcal{A}, \bullet] ]]/3! + \cdots$,
we obtain $e^{\mathcal{A}} N^\times e^{- \mathcal{A}} = N^\times$ and  $e^{\mathcal{A}} N^\circ e^{- \mathcal{A}} = N^\circ + 2 \mathcal{A}$.
These relations yield
\begin{gather*}
    e^{\mathcal{A}} \mathcal{D}[a] e^{- \mathcal{A}}
    = e^{\mathcal{A}} \Big( \mathcal{A} - \frac{1}{2}N^\circ \Big) e^{- \mathcal{A}}
    = \mathcal{A} - \frac{1}{2} (N^\circ + 2 \mathcal{A})
    = - \frac{1}{2} N^\circ.
\end{gather*}
From $(N^2)^\times = N^\circ N^\times$,
we obtain $(a^\dagger \! a^\dagger \! a a)^\times = (N^2 - N)^\times = (N^\circ - \mathcal{I}) N^\times$ and
\begin{gather*}
    e^{\mathcal{A}} (a^\dagger \! a^\dagger a a)^\times e^{- \mathcal{A}}
    = e^{\mathcal{A}}  (N^\circ - \mathcal{I}) N^\times e^{- \mathcal{A}}
    = 2 N^\times \mathcal{A} + (N^\circ - \mathcal{I}) N^\times.
\end{gather*}
From $(N^2)^\circ = ( (N^\circ)^2 + (N^\times)^2 )/2$,
we obtain $(a^\dagger \! a^\dagger \! a a)^\circ = (N^2 - N)^\circ = ((N^\circ)^2 + (N^\times)^2)/2 - N^\circ$ and
\begin{equation*}
    \begin{gathered}
        e^{\mathcal{A}} \mathcal{D}[a^2] e^{- \mathcal{A}}
        = e^{\mathcal{A}} \Big( \mathcal{A}^2 - \frac{1}{4} \{ (N^\circ)^2 + (N^\times)^2 \} + \frac{1}{2} N^\circ \Big) e^{- \mathcal{A}} \\
        = \mathcal{A}^2 - \frac{1}{4} ( (N^\circ + 2 \mathcal{A})^2 + (N^\times)^2 ) + \frac{1}{2} (N^\circ + 2 \mathcal{A})\\
        = 2 \mathcal{A} - \mathcal{A} N^\circ - N^\circ \mathcal{A} - \frac{1}{2} (N^2 - N)^\circ \\
        = \mathcal{A} ( 2 \mathcal{I} - N^\circ ) - \frac{1}{2} (N^2 - N)^\circ.
    \end{gathered}
\end{equation*}
Combining these results, \eref{eq:diag_eALe-A} can be derived.

\section{Eigenvectors with confluent hypergeometric function}
\label{app:eigvec.F.}

In this appendix, we rewrite the right (left) eigenvector \eref{eq:diag_Reigvec_0} (\eref{eq:diag_Leigvec_0}) using the confluent hypergeometric function as \eref{eq:diag_Reigvec_1} (\eref{eq:diag_Leigvec_1}).
Let us start with the left eigenvector.
We have $\prod_{q=k+1}^{p} (z+q) = (z+k+1)_{p-k}$ and $\prod_{q=k+1}^{p} z = z^{p-k}$ for $p \geq k+1$.
In addition, we find $|m|/2 + iUm/(2 \kappa_2) = x_k^{(m)} - k$.
With these relations, the product inside the square bracket in \eref{eq:diag_Leigvec_0} can be written with the rising factorials as
\begin{equation*}
    \begin{gathered}
        \prod_{q=k+1}^{p} \frac{c_q^{(m)}}{\lambda_k^{(m)} - \lambda_q^{(m)}}
        = \prod_{q=k+1}^{p} \frac{(-2) \sqrt{q} \sqrt{|m|+q} \, (x_k^{(m)} - k - 1 + q)}{(-k + q)( 2 x_k^{(m)} + \kappa_1/\kappa_2 - k - 1 + q)} \\
        = \frac{(-2)^{p-k} \sqrt{(k+1)_{p-k}} \sqrt{(k+|m|+1)_{p-k}} (x_k^{(m)})_{p-k} }{ (1)_{p-k} (2x_k^{(m)} + \kappa_1/\kappa_2)_{p-k}} \\
        = \frac{(-2)^{p-k}}{(p-k)!} \frac{(x_k^{(m)})_{p-k}}{(2x_k^{(m)} + \kappa_1/\kappa_2)_{p-k}} \sqrt{\frac{p!(p+|m|)!}{k!(k+|m|)!}},
    \end{gathered}
\end{equation*}
where, in the last equality, we have used $(k+1)_n = (k+n)!/k! \ (k \in \mathbb{Z}_{\geq 0})$.
Using $\dbra{\phi_k^{(m)}} \hat{\mathcal{A}}^{p-k} = \sqrt{p!(p+|m|)!/(k!(k+|m|)!)} \dbra{\phi_p^{(m)}} \ (p \geq k)$, this result yields
\begin{equation*}
    \Big[ \prod_{q=k+1}^{p} \frac{c_q^{(m)}}{\lambda_k^{(m)} - \lambda_q^{(m)}} \Big] \dbra{\phi_p^{(m)}}
    = \frac{(x_k^{(m)})_{p-k}}{(2x_k^{(m)} + \kappa_1/\kappa_2)_{p-k}} \dbra{\phi_k^{(m)}} \frac{(-2 \hat{\mathcal{A}})^{p-k}}{(p-k)!}.
\end{equation*}
Thus, we obtain
\begin{equation*}
    \begin{gathered}
        \dbra{\bar{\rho}_k^{(m)}} e^{- \hat{\mathcal{A}}}
        = \dbra{\phi_k^{(m)}} \Big[ \hat{\mathcal{I}} + \sum_{p = k+1}^{\infty} \frac{(x_k^{(m)})_{p-k}}{(2x_k^{(m)} + \kappa_1/\kappa_2)_{p-k}}  \frac{(-2 \hat{\mathcal{A}})^{p-k}}{(p-k)!} \Big] \\
        = \dbra{\phi_k^{(m)}} \sum_{j=0}^\infty \frac{(x_k^{(m)})_j}{(2x_k^{(m)} + \kappa_1/\kappa_2)_j}  \frac{(-2 \hat{\mathcal{A}})^j}{j!} \\
        = \dbra{\phi_k^{(m)}} {}_1 F_1(x_k^{(m)}; 2x_k^{(m)} + \frac{\kappa_1}{\kappa_2}; - 2 \hat{\mathcal{A}} ).
    \end{gathered}
\end{equation*}

For the right eigenvectors, we use
$\prod_{q=p}^{k-1} (z+q) = (z+p)_{k-p}$ and $\prod_{q=p}^{k-1} z = z^{k-p}$ for $k-1 \geq p$;
\begin{equation*}
    \begin{gathered}
        \prod_{q=p}^{k-1} \frac{c_{q+1}^{(m)}}{\lambda_k^{(m)} - \lambda_q^{(m)}}
        = \prod_{q=p}^{k-1} \frac{ (-2) \sqrt{1+q} \sqrt{|m|+1+q} \, (x_k^{(m)} - k + q)}{(- k + q)( 2 x_k^{(m)} + \kappa_1/\kappa_2 - 1 - k + q)} \\
        = \frac{2^{k-p} \sqrt{(p+1)_{k-p}} \sqrt{(p+|m|+1)_{k-p}} \, (x_k^{(m)} - k + p)_{k-p} }{ (k-p)_{k-p} (2x_k^{(m)} + \kappa_1/\kappa_2 -1 - k + p)_{k-p}} \\
        = \frac{2^{k-p}}{(k-p)!} \frac{(x_k^{(m)}-k+p)_{k-p}}{(2x_k^{(m)} + \kappa_1/\kappa_2 - 1 - k + p)_{k-p}} \sqrt{\frac{k!(k+|m|)!}{p!(p+|m|)!}}.
    \end{gathered}
\end{equation*}
Using $\hat{\mathcal{A}}^{k-p} \dket{\phi_k^{(m)}} = \sqrt{k!(k+|m|)!/(p!(p+|m|)!)} \dket{\phi_p^{(m)}}$ $(k \geq p)$, this yields
\begin{equation*}
    \begin{gathered}
          \Big[ \prod_{q=p}^{k-1} \frac{c_{q+1}^{(m)}}{\lambda_k^{(m)} - \lambda_q^{(m)}} \Big] \dket{\phi_p^{(m)}}
          = \frac{(x_k^{(m)}-k+p)_{k-p}}{(2x_k^{(m)} + \kappa_1/\kappa_2 - 1 - k + p)_{k-p}} \frac{(2 \hat{\mathcal{A}})^{k-p}}{(k-p)!} \dket{\phi_k^{(m)}},
    \end{gathered}
\end{equation*}
and
\begin{gather}
    e^{\hat{\mathcal{A}}} \dket{\rho_k^{(m)}}
    = \sum_{p = 0}^{k} \frac{(x_k^{(m)}-k+p)_{k-p}}{(2x_k^{(m)} + \kappa_1/\kappa_2 - 1 - k + p)_{k-p}} \frac{(2 \hat{\mathcal{A}})^{k-p}}{(k-p)!} \dket{\phi_k^{(m)}} \nonumber \\
    = \sum_{j = 0}^{k} \frac{(x_k^{(m)} - j)_j}{(2x_k^{(m)} + \kappa_1/\kappa_2 - 1 - j)_j} \frac{(2 \hat{\mathcal{A}})^{j}}{j!} \dket{\phi_k^{(m)}} \nonumber \\
    = \sum_{j = 0}^{k} \frac{(1 - x_k^{(m)})_j}{(2 - \{ 2x_k^{(m)} + \kappa_1/\kappa_2 \})_j} \frac{(2 \hat{\mathcal{A}})^{j}}{j!} \dket{\phi_k^{(m)}}, \label{eq:eigvec.F._Reigvec_FiniteSum}
\end{gather}
where, in the last equality, we have used $(z-j)_j = (-1)^j (1-z)_j$.
Note that the denominators $(2 - \{ 2x_k^{(m)} + \kappa_1/\kappa_2 \})_j \ (j \in \mathbb{Z}_{[0,k]})$ are nonzero if $\lambda_k^{(m)} - \lambda_q^{(m)} \ne 0 \ (k \ne q)$.
To further simplify the form, let us assume $2 - ( 2x_k^{(m)} + \kappa_1/\kappa_2 ) \notin \mathbb{Z}_{\leq 0}$ for $m \in \mathbb{Z}$ and $k \in \mathbb{Z}_{\geq 0}$.
One can easily show that this is equivalent to assuming $\kappa_1/\kappa_2 \notin \mathbb{Z}$.
Under this assumption, the upper limit of the sum in \eref{eq:eigvec.F._Reigvec_FiniteSum} can be extended to $\infty$ because of the relation $\hat{\mathcal{A}}^j \dket{\phi_k^{(m)}} = 0 \ (j > k)$. We thus obtain
\begin{gather*}
    e^{\hat{\mathcal{A}}} \dket{\rho_k^{(m)}}
    = \sum_{j = 0}^{\infty} \frac{(1 - x_k^{(m)})_j}{(2 - \{ 2x_k^{(m)} + \kappa_1/\kappa_2 \})_j} \frac{(2 \hat{\mathcal{A}})^{j}}{j!} \dket{\phi_k^{(m)}} \\
    = {}_1 F_1 \Big( 1-x_k^{(m)};2 - \Big\{ 2x_k^{(m)} + \frac{\kappa_1}{\kappa_2} \Big\} ; 2 \hat{\mathcal{A}} \Big) \dket{\phi_k^{(m)}}.
\end{gather*}

\section{Proof of \eref{eq:diag_F-1}}
\label{app:F-1}

Let $\mathcal{G} = {}_1 F_1^R ( \mathcal{I} - \mathcal{X}; 2 \mathcal{I} - \{ 2 \mathcal{X} + (\kappa_1/\kappa_2) \mathcal{I} \} ; 2 \mathcal{A} )$ and
$\mathcal{F} = {}_1 F_1^L ( \mathcal{X};2 \mathcal{X} + (\kappa_1/\kappa_2) \mathcal{I} ; - 2 \mathcal{A} )$
with ${}_1 F_1^R$ and ${}_1 F_1^L$ the confluent geometric function accounting for the non-commutative character as defined in \erefs{eq:diag_Reigvec} and (\ref{eq:diag_Leigvec}).
Here we prove $\mathcal{F} \mathcal{G} = \mathcal{G} \mathcal{F} = \mathcal{I}$, or equivalently $\mathcal{G}$ is the inverse of $\mathcal{F}$, $\mathcal{G} = \mathcal{F}^{-1}$.

The proof utilizes the following identity; given a function of $N^\circ$, $f(N^\circ)$, we have
\begin{equation}
  f(N^\circ) \mathcal{A}^n = \mathcal{A}^n f(N^\circ-2n\mathcal{I}) \ \ (n \in \mathbb{Z}_{\geq 0}).
  \label{eq:eigvec.F._N^cA_-}
\end{equation}
To show this, we note $\hat{\mathcal{A}}^n \dket{\phi_{k}^{(m)}} = c_{k,n}^{(m)} \dket{\phi_{k-n}^{(m)}}$ and $c_{k,n}^{(m)} \dbra{\phi_{k}^{(m)}} = \dbra{\phi_{k-n}^{(m)}} \hat{\mathcal{A}}^n$
with $n \in \mathbb{N}_{\geq 0}$ and $c_{k,n}^{(m)} = n! \sqrt{\binom{|m|+k+n}{n} \binom{k+n}{n}}$.
These relations yield
\begin{equation*}
  \hat{\mathcal{A}}^n \dketbra{\phi_k^{(m)}}{\phi_k^{(m)}} = \dketbra{\phi_{k-n}^{(m)}}{\phi_{k- n}^{(m)}} \hat{\mathcal{A}}^n,
\end{equation*}
where $\dket{\phi_{k<0}^{(m)}} = 0$.
Then, with the completeness relation of the basis $\{ \dket{\phi_k^{(m)}} \}_{m \in \mathbb{Z}, k \in \mathbb{Z}_{\geq 0}}$, $\hat{\mathcal{I}} = \sum_{m=-\infty}^{\infty} \sum_{k=0}^\infty \dketbra{\phi_k^{(m)}}{\phi_k^{(m)}}$,
and the eigenvalue relation $\dbra{\phi_k^{(m)}} \hat{N}^\circ = (2k+|m|) \dbra{\phi_k^{(m)}}$,
the right hand side of \eref{eq:eigvec.F._N^cA_-} reads as
\begin{gather*}
        \hat{\mathcal{A}}^n f(\hat{N}^\circ - 2n \hat{\mathcal{I}})
        = \hat{\mathcal{A}}^n \, \hat{\mathcal{I}} \, f(\hat{N}^\circ - 2n \hat{\mathcal{I}}) \\
        = \hat{\mathcal{A}}^n \Big\{ \sum_{m=-\infty}^{\infty} \sum_{k=0}^\infty f(|m| + 2 (k-n) ) \dketbra{\phi_k^{(m)}}{\phi_k^{(m)}} \Big\} \\
        = \Big\{ \sum_{m=-\infty}^{\infty} \sum_{k=n}^\infty f(|m| + 2 (k-n) ) \dketbra{\phi_{k-n}^{(m)}}{\phi_{k-n}^{(m)}} \Big\} \hat{\mathcal{A}}^n \\
        = f(\hat{N}^\circ) \Big\{ \sum_{m=-\infty}^{\infty} \sum_{k=0}^\infty \dketbra{\phi_{k}^{(m)}}{\phi_{k}^{(m)}} \Big\} (\hat{\mathcal{A}})^n
        = f(\hat{N}^\circ) \hat{\mathcal{A}}^n.
\end{gather*}

Using \eref{eq:eigvec.F._N^cA_-}, we first show $\mathcal{F} \mathcal{G} = \mathcal{I}$.
Inserting the definitions yields (we denote $\eta = \kappa_1/\kappa_2$)
\begin{gather}
    \mathcal{F} \mathcal{G}
    = \Big\{ \sum_{p = 0}^\infty \frac{(\mathcal{X})_p}{ ( 2 \mathcal{X} + \eta \mathcal{I} )_p } \frac{ ( - 2 \mathcal{A})^p}{p!} \Big\} \Big\{ \sum_{q = 0}^\infty \frac{ (2 \mathcal{A})^q}{q!} \frac{(\mathcal{I}- \mathcal{X})_q}{ ( 2 \mathcal{I} - \{ 2 \mathcal{X} + \eta \mathcal{I} \} )_q } \Big\} \nonumber  \\
    = \sum_{p,q=0}^\infty \frac{(-1)^{q} (-2 \mathcal{A})^{p+q} }{p!q!} \frac{(\mathcal{X} - (p+q) \mathcal{I} )_p}{ ( 2 \mathcal{X} + (\eta - 2(p + q)) \mathcal{I} )_p} \frac{(\mathcal{I} - \mathcal{X})_q}{ ( 2 \mathcal{I} - \{ 2 \mathcal{X} + \eta \mathcal{I} \} )_q } \nonumber \\
    \begin{gathered}
        = \mathcal{I} + \sum_{n=1}^\infty \frac{(-2 \mathcal{A} )^n}{n!} \Big[ \sum_{k=0}^n \binom{n}{k} (-1)^k \frac{(\mathcal{X} - n \mathcal{I} )_{n-k} (\mathcal{I} - \mathcal{X})_k  }{ ( 2 \mathcal{X} + (\eta - 2n) \mathcal{I} )_{n-k} ( 2 \mathcal{I} - \{ 2 \mathcal{X} + \eta \mathcal{I} \} )_k } \Big].
    \end{gathered}
    \label{eq:F-1_FG}
\end{gather}
We show that the finite sum inside the square bracket vanishes for $n \in \mathbb{Z}_{\geq 1}$.
From $(z)_n = (-1)^k (z)_{n-k} (1-n-z)_{n-k} \ (n \geq k)$, we have
$(\mathcal{X} - n\mathcal{I})_n = (-1)^k (\mathcal{X} - n\mathcal{I})_{n-k} (\mathcal{I} - \mathcal{X})_k$ and
$(2\mathcal{X} + (\eta - 2n) \mathcal{I})_n = (-1)^k (2\mathcal{X} + (\eta - 2n) \mathcal{I})_{n-k} ( (n+1)\mathcal{I} - \{ 2\mathcal{X} + \eta \mathcal{I} \} )_k$.
Then,
\begin{gather*}
  \frac{(\mathcal{X} - n \mathcal{I} )_{n-k} (\mathcal{I} - \mathcal{X})_k  }{ ( 2 \mathcal{X} + (\eta - 2n) \mathcal{I} )_{n-k} ( 2 \mathcal{I} - \{ 2 \mathcal{X} + \eta \mathcal{I} \} )_k }
  = \frac{(\mathcal{X} - n\mathcal{I})_n}{(2\mathcal{X} + (\eta - 2n) \mathcal{I})_n} \frac{ ( (n+1)\mathcal{I} - \{ 2\mathcal{X} + \eta \mathcal{I} \} )_k}{( 2 \mathcal{I} - \{ 2 \mathcal{X} + \eta \mathcal{I} \} )_k}.
\end{gather*}
Furthermore, from $(n-1+z)_k/(z)_k = (k+z)_{n-1}/(z)_{n-1}$, we have
\begin{gather*}
    \frac{ ( (n+1)\mathcal{I} - \{ 2\mathcal{X} + \eta \mathcal{I} \} )_k}{( 2 \mathcal{I} - \{ 2 \mathcal{X} + \eta \mathcal{I} \} )_k}
    = \frac{ ( (n-1)\mathcal{I} + 2\mathcal{I} - \{ 2\mathcal{X} + \eta \mathcal{I} \} )_k}{( 2 \mathcal{I} - \{ 2 \mathcal{X} + \eta \mathcal{I} \} )_k}
    = \frac{ ( (k + 2 - \eta)\mathcal{I} - 2\mathcal{X} )_{n-1}}{ ( (2-\eta) \mathcal{I} - 2 \mathcal{X} )_{n-1}}
\end{gather*}
Combining these results, the sum inside the square bracket in \eref{eq:F-1_FG} reads as
\begin{gather*}
  \frac{(\mathcal{X} - n\mathcal{I})_n}{(2\mathcal{X} + (\eta - 2n) \mathcal{I})_n ( (2-\eta) \mathcal{I} - 2 \mathcal{X} )_{n-1}} \sum_{k=0}^n \binom{n}{k} (-1)^k ( (k + 2 - \eta)\mathcal{I} - 2\mathcal{X} )_{n-1}.
\end{gather*}
Note that $( (k + 2 - \eta)\mathcal{I} - 2\mathcal{X} )_{n-1}$ is a polynomial of degree $(n-1)$ in $k$.
Accordingly, we have
\begin{equation*}
  \sum_{k=0}^n \binom{n}{k} (-1)^k ( (k + 2 - \eta)\mathcal{I} - 2\mathcal{X} )_{n-1} = 0,
\end{equation*}
due to the identity $\sum_{k=0}^n \binom{n}{k} (-1)^k k^l = 0 \ (l \in \mathbb{Z}_{[0,n-1]})$, which can be shown using the binomial theorem $(1-x)^n = \sum_{k=0}^n \binom{n}{k} (-x)^k$ as follows;
\begin{gather*}
    \sum_{k=0}^n \binom{n}{k} (-1)^k k^l
    = \left. \Big[ \Big( x \frac{d}{dx} \Big)^l \sum_{k=0}^n \binom{n}{k} (-1)^k x^k \Big] \right|_{x=1} 
    = \left. \Big[ \Big( x \frac{d}{dx} \Big)^l (1-x)^n \Big] \right|_{x=1}
    = 0 \ \  (l \in \mathbb{Z}_{[0,n-1]}).
\end{gather*}
Hence, the sum inside the square bracket in \eref{eq:F-1_FG} is equal to zero, thereby establishing the proof of $\mathcal{F} \mathcal{G} = \mathcal{I}$.

We next show $\mathcal{G} \mathcal{F} = \mathcal{I}$.
This time, inserting the definitions yields
\begin{gather}
    \mathcal{G} \mathcal{F}
    = \Big\{ \sum_{q = 0}^\infty \frac{ (2 \mathcal{A})^q}{q!} \frac{(\mathcal{I}- \mathcal{X})_q}{ ( 2 \mathcal{I} - \{ 2 \mathcal{X} + \eta \mathcal{I} \} )_q } \Big\} \Big\{ \sum_{p = 0}^\infty \frac{(\mathcal{X})_p}{ ( 2 \mathcal{X} + \eta \mathcal{I} )_p } \frac{ ( - 2 \mathcal{A})^p}{p!} \Big\} \nonumber \\
    = \sum_{p,q = 0}^\infty \frac{((1-q)\mathcal{I}- \mathcal{X})_q} { ( 2 \mathcal{I} - \{ 2 \mathcal{X} + (\eta + 2q) \mathcal{I} \} )_q } \frac{(\mathcal{X} + q\mathcal{I})_p}{ ( 2 \mathcal{X} + (\eta+2q) \mathcal{I} )_p } \frac{ (-1)^q ( - 2 \mathcal{A})^{p+q}}{p!q!} \nonumber \\
    \begin{gathered}
        = \mathcal{I} + \sum_{n=1}^\infty \frac{(-2 \mathcal{A})^{n} }{n!} \Big[ \sum_{k=0}^n \binom{n}{k} (-1)^k \frac{ ((1-k)\mathcal{I}- \mathcal{X})_k (\mathcal{X} + k\mathcal{I})_{n-k}}{ (2 \mathcal{I} - \{ 2 \mathcal{X} + (\eta + 2k) \mathcal{I} \} )_k ( 2 \mathcal{X} + (\eta+2k) \mathcal{I} )_{n-k} } \Big].
    \end{gathered}
    \label{eq:F-1_GF}
\end{gather}
We show that the finite sum inside the square bracket vanishes for $n \in \mathbb{Z}_{\geq 1}$.
From $(z)_n = (-1)^k (1-k-z)_k (z+k)_{n-k} \ (n \geq k)$ and $(z)_k = (-1)^k (1 - k - z)_n$, we have
$(\mathcal{X})_n = (-1)^k ((1-k)\mathcal{I}-\mathcal{X})_k (\mathcal{X}+k\mathcal{I})_{n-k}$ and $(2 \mathcal{I} - \{ 2 \mathcal{X} + (\eta + 2k) \mathcal{I} \} )_k = (-1)^k (\{ 2\mathcal{X} + (\eta-1) \mathcal{I} \} + k \mathcal{I} )_k$, respectively.
Then,
\begin{gather*}
    \frac{ (\mathcal{X} + k\mathcal{I})_{n-k} ((1-k)\mathcal{I}- \mathcal{X})_k}{ ( 2 \mathcal{X} + (\eta+2k) \mathcal{I} )_{n-k} (2 \mathcal{I} - \{ 2 \mathcal{X} + (\eta + 2k) \mathcal{I} \} )_k } \\
    = \frac{(\mathcal{X})_n}{ (\{ 2\mathcal{X} + (\eta-1) \mathcal{I} \} + k \mathcal{I} )_k } \frac{1}{( \{ 2 \mathcal{X} + (\eta-1)\mathcal{I} \} + (2k+1) \mathcal{I} )_{n-k}}.
\end{gather*}
Using
\begin{gather*}
    [(z+k)_{k} \times (z+2k+1)_{n-k} ]^{-1} \\
    = [ (z+k)(z+k+1) \cdots (z+2k-1)  (z+2k+1) (z+2k+2) \cdots (z+n+k) ]^{-1} \\
    = \frac{z+2k}{(z+k)_{n+1}},
\end{gather*}
we have
\begin{gather*}
  \frac{ (\mathcal{X} + k\mathcal{I})_{n-k} ((1-k)\mathcal{I}- \mathcal{X})_k}{ ( 2 \mathcal{X} + (\eta+2k) \mathcal{I} )_{n-k} (2 \mathcal{I} - \{ 2 \mathcal{X} + (\eta + 2k) \mathcal{I} \} )_k }
  = (\mathcal{X})_n \frac{\{ 2 \mathcal{X} + (\eta-1)\mathcal{I} \} + 2k \mathcal{I} }{ (\{ 2\mathcal{X} + (\eta-1) \mathcal{I} \} + k \mathcal{I} )_{n+1} }.
\end{gather*}
Combining these results, the sum inside the square bracket in \eref{eq:F-1_GF} reads
\begin{equation*}
  (\mathcal{X})_n
  \sum_{k=0}^n \binom{n}{k} (-1)^k \frac{\{ 2 \mathcal{X} + (\eta-1)\mathcal{I} \} + 2k \mathcal{I} }{ (\{ 2\mathcal{X} + (\eta-1) \mathcal{I} \} + k \mathcal{I} )_{n+1} }.
\end{equation*}
To prove that this is equal to zero, we now show $\sum_{k=0}^n \binom{n}{k} (-1)^k (z+2k)/(z+k)_{n+1} = 0 \ (n \in \mathbb{N}_{\geq 1}, z \ne 0, -1, \dots -2n)$.
The partial fraction decomposition yields
\begin{equation*}
  \frac{z+2k}{(z+k)_{n+1}} = \sum_{j=0}^n \frac{c_{n,k,j}}{z+k+j},
\end{equation*}
where the coefficients $\{ c_{n,k,j} \}_{k,j \in \mathbb{Z}_{[0,n]}}$ can be determined as follows;
\begin{gather*}
    c_{n,k,j}
    = \left. \Big[ (z+k+j) \Big\{ \sum_{q=0}^n \frac{c_{n,k,q}}{z+k+q} \Big\}  \Big] \right|_{z = - (k+j)} \\
    = \left. \Big[ (z+k+j) \frac{z+2k}{(z+k)_{n+1}} \Big] \right|_{z = - (k+j)} \\
    = \frac{2k-(k+j)}{ (-j) (-j+1) \cdots (-1) \cdot 1 \cdot 2 \cdots (n-j)}
    = \frac{(-1)^j}{n!} \binom{n}{j} (k-j).
\end{gather*}
Note that this expression is true even when $j = 0$ and $j = n$ with the convention $0! = 1$.
Accordingly,
\begin{gather*}
    \sum_{k=0}^n \binom{n}{k} (-1)^k \frac{z+2k}{(z+k)_{n+1}}
    = \sum_{k=0}^n \binom{n}{k} (-1)^k \sum_{j=0}^n \frac{(-1)^j}{n!} \binom{n}{j} \frac{k-j}{z+k+j} \\
    = \frac{1}{n!} \sum_{j,k=0}^n \binom{n}{k} \binom{n}{j} (-1)^{j+k} \frac{k-j}{z+k+j}
    = 0,
\end{gather*}
where the last equality follows from the anti-symmetry of the summand with respect to the exchange of $j$ and $k$.
Therefore, the sum inside the square bracket in \eref{eq:F-1_GF} is equal to zero, and the proof of $\mathcal{G} \mathcal{F} = \mathcal{I}$ is completed.

\section{Explicit forms of eigenvectors with $k = 0$ and $1$}
\label{app:exp.eigvec.}

In this appendix, we derive the explicit forms of several eigenvectors in the cases $k = 0$ and $1$.
When $\kappa_1 > 0$, the right eigenvectors are given by \eref{eq:diag_Reigvec_FiniteSum}.
For small values of $k$, they can be calculated easily.
For instance, we obtain
\begin{equation}
  \rho_0^{(m)} = \phi_0^{(m)},
  \label{eq:exp.eigvec._R_0^m}
\end{equation}
when $k = 0$ and
\begin{equation*}
  \rho_1^{(m)} = \phi_1^{(m)} + \frac{(\kappa_1/\kappa_2)\sqrt{|m|+1}}{2 - (2x_1^{(m)} + \kappa_1/\kappa_2)} \phi_0^{(m)},
\end{equation*}
when $k = 1$,
where we have used $e^{-\hat{\mathcal{A}}} \dket{\phi_0^{(0)}} = \dket{\phi_0^{(0)}}$ and $e^{-\hat{\mathcal{A}}} \dket{\phi_1^{(0)}} = \dket{\phi_1^{(0)}} - \dket{\phi_0^{(0)}}$.

For the left eigenvectors, we utilize the relation $\dbra{\phi_0^{(0)}} e^{\hat{\mathcal{A}}} = \dbra{I}$ with the identity operator $I$, which can be shown as
\begin{gather}
    \dbra{\phi_0^{(0)}} e^{\hat{\mathcal{A}}} 
    = \Big[ e^{\hat{\mathcal{A}}^\dagger} \dket{\ketbra{0}{0}}  \Big]^\dagger
    = \dbra{ \sum_{n = 0}^\infty \frac{1}{n!} (a^\dagger)^n \ketbra{0}{0} a^n }
    = \dbra{ \sum_{n = 0}^\infty \ketbra{n}{n} } \nonumber \\
    = \dbra{ I } \label{eq:exp.eigvec._L_0^0}
\end{gather}
When $\kappa_1 > 0$, the left eigenvectors are given by \eref{eq:diag_Leigvec_1}.
When $k = 0$, using $\phi_0^{(m \geq 0)} = (a^\dagger)^m \phi_0^{(0)} / \sqrt{m!}$, we obtain
\begin{gather*}
    \dket{\bar{\rho}_0^{(m \geq 0)}} 
    = \Big[ \dbra{\phi_0^{(m)}} e^{\hat{\mathcal{A}}} {}_1 F_1(x_0^{(m)}; 2x_0^{(m)} + \kappa_1/\kappa_2; - 2 \hat{\mathcal{A}} ) \Big]^\dagger  \\
    = \frac{1}{\sqrt{m!}} {}_1 F_1(x_0^{(-m)}; 2x_0^{(-m)} + \kappa_1/\kappa_2; - 2 \hat{\mathcal{A}}^\dagger ) \dket{(a^\dagger)^m} \\
    = \frac{1}{\sqrt{m!}} \sum_{j=0}^\infty \frac{(x_0^{(-m)})_j (-2)^j}{(2x_0^{(-m)} + \kappa_1/\kappa_2)_j j!} \dket{ (a^\dagger)^{j+m} a^j  },
\end{gather*}
which yields
\begin{gather}
    \bar{\rho}_0^{(m \geq 0)} 
    = \frac{1}{\sqrt{m!}} \sum_{j=0}^\infty \frac{(x_0^{(-m)})_j (-2)^j}{(2x_0^{(-m)} + \kappa_1/\kappa_2)_j j!} (a^\dagger)^{j+m} a^j \nonumber \\
    = \frac{1}{\sqrt{m!}} (a^\dagger)^m : {}_1 F_1(x_0^{(-m)}; 2x_0^{(-m)} + \kappa_1/\kappa_2; - 2 N) :, \label{eq:exp.eigvec._Leigvec0m}
\end{gather}
where $: :$ denotes the normal ordering operation. $\rho_0^{(m<0)}$ can be derived using the identity $(\rho_0^{(m)})^\dagger = (\rho_0^{(-m)})^\dagger$.

If we assume $\kappa_1 = U = 0$, \eref{eq:exp.eigvec._Leigvec0m} can be further simplified as follows.
Note that, for $|m| > 0$, \eref{eq:exp.eigvec._Leigvec0m} remains valid even when $\kappa_1 = 0$.
By setting $\kappa_1 = U = 0$, we obtain
\begin{gather*}
    \bar{\rho}_0^{(m > 0)}
    = \frac{1}{\sqrt{m!}} (a^\dagger)^m : {}_1 F_1 \Big( \frac{m}{2} ; m ; - 2 N \Big) : \\
    = \frac{1}{\sqrt{m!}} (a^\dagger)^m : {}_0 F_1 \Big( ; \frac{m+1}{2} ; \frac{N^2}{4} \Big) e^{-N} :,
\end{gather*}
where we have used the identity $ {}_1 F_1( \beta/2 ; \beta ; z) = {}_0 F_1( ; (\beta+1)/2 ; z^2/16) e^{z/2} \ (\beta \notin \mathbb{Z}_{\leq 0})$ with ${}_0 F_1( ; \gamma ; z) = \sum_{j=0}^\infty z^j/( (\gamma)_j j!) \ (\gamma \in \mathbb{C}, \gamma \notin \mathbb{Z}_{\leq 0})$.
Inserting the definition of ${}_0 F_1$ yields
\begin{gather}
    \bar{\rho}_0^{(m > 0)}
    = \frac{1}{\sqrt{m!}} (a^\dagger)^m : \sum_{j=0}^\infty \frac{1}{((m+1)/2)_j} \frac{N^{2j}}{j! 2^{2j}} e^{-N} : \nonumber \\
    = \frac{(m-1)!!}{\sqrt{m!}} (a^\dagger)^m : \sum_{j=0}^\infty \frac{1}{(m+2j-1)!!} \frac{(2j)!}{2^j j!}  \frac{N^{2j}}{(2j)!} e^{-N} : \nonumber \\
    = \frac{(m-1)!!}{\sqrt{m!}} (a^\dagger)^m : \sum_{j=0}^\infty \frac{(2j-1)!!}{(m+2j-1)!!} \frac{N^{2j}}{(2j)!} e^{-N} : \nonumber \\
    = \frac{(m-1)!!}{\sqrt{m!}} (a^\dagger)^m \Pi_+ \frac{(N-1)!!}{(m+N-1)!!},   \label{eq:exp.eigvec._L_0^m}
\end{gather}
with $k!! = k \times (k-2)!! \ (k \in \mathbb{Z}_{\geq -1})$ the double factorial ($0!! = (-1)!! \stackrel{\text{def}}{=} 1$) and $\Pi_+ = \sum_{n = 0}^\infty \ketbra{2n}{2n}$ the parity operator.
In the derivation, we have used $(k/2)_j = (k+2j-2)!!/(2^j(k-2)!!) (k \in \mathbb{Z}_{\geq 1})$ in the second equality and $(2j)!/(j!2^j) = (2j-1)!! \ (j \in \mathbb{Z}_{\geq 0})$ in the third equality.
The last equality follows from the identity
\begin{equation*}
  \Pi_+ f(N) = : \sum_{j=0}^\infty f(2j) \frac{N^{2j}}{(2j)!} e^{-N} :,
\end{equation*}
with $f(N)$ a function of $N$, which can be derived as
\begin{gather*}
    \dket{\Pi_+ f(N)}
    = \sum_{j=0}^\infty f(2j) \frac{(\hat{\mathcal{A}}^\dagger)^{2j}}{(2j)!} \dket{ \phi_0^{(0)} } \\
    = \sum_{j=0}^\infty f(2j) \frac{(\hat{\mathcal{A}}^\dagger)^{2j}}{(2j)!} e^{-\hat{\mathcal{A}}^\dagger} \dket{ I } 
    = \sum_{j=0}^\infty f(2j) \dket{ : \frac{N^{2j}}{(2j)!} e^{-N} : }.
\end{gather*}

Lastly, for the left eigenvectors with $k=1$, we can obtain a simpler form for $m = 0$;
\begin{gather*}
    \dbra{\phi_1^{(0)}} e^{\hat{\mathcal{A}}} {}_1 F_1(x_1^{(0)}; 2x_1^{(0)} + \kappa_1/\kappa_2; - 2 \hat{\mathcal{A}} ) \\
    = - \frac{1}{2} \dbra{\phi_0^{(0)}} e^{\hat{\mathcal{A}}} (-2 \hat{\mathcal{A}}) {}_1 F_1(1; 2 + \kappa_1/\kappa_2; - 2 \hat{\mathcal{A}} ) \\
    = \frac{1}{2} \dbra{\phi_0^{(0)}} e^{\hat{\mathcal{A}}} \Big\{ {}_1 F_1(0; 1+\kappa_1/\kappa_2; - 2 \hat{\mathcal{A}} ) - {}_1 F_1(1; 1+\kappa_1/\kappa_2; - 2 \hat{\mathcal{A}} ) \Big\} \\
    = \frac{1}{2} \dbra{\phi_0^{(0)}} e^{\hat{\mathcal{A}}} \Big\{ \hat{\mathcal{I}} - {}_1 F_1(1; 1+\kappa_1/\kappa_2; - 2 \hat{\mathcal{A}} ) \Big\},
\end{gather*}
where, in the second equality, we have used the contiguous relation $(z/\beta) {}_1 F_1 (\alpha+1,\beta+1,z) = {}_1 F_1(\alpha+1,\beta,z) - {}_1 F_1(\alpha,\beta,z)$.
When $\kappa_1 = 0$, this reads
\begin{gather}
    \dbra{\phi_1^{(0)}} e^{\hat{\mathcal{A}}} {}_1 F_1(1; 2; - 2 \hat{\mathcal{A}} ) \nonumber \\
    = \frac{1}{2} \dbra{\phi_0^{(0)}} e^{\hat{\mathcal{A}}} \Big\{ \hat{\mathcal{I}} - {}_1 F_1(1; 1; - 2 \hat{\mathcal{A}} ) \Big\}
    = \frac{1}{2} \dbra{\phi_0^{(0)}} (e^{\hat{\mathcal{A}}} - e^{-\hat{\mathcal{A}}}) \nonumber \\
    = \dbra{I - \Pi_+}. \label{eq:exp.eigvec._L_1^0}
\end{gather}

\section{Exact time evolution}
\label{app:time.evolv.}

In this appendix, we explore the exact solution of the master equaiton.
We first demonstrate that our findings in this article provide consistent results with those in \rcite{Simaan78}, in which the authors found the exact solution of the master equation $(d/dt) \rho(t) = \kappa_2 \mathcal{D}[a^2] \rho(t)$.
After sorting out, their results can be represented as $(m \geq 0)$ 
\begin{gather*}
    \dbraket{\phi_k^{(m)}}{\rho(t)} = \sqrt{\frac{k!}{(k+m)!}} \sum_{b = 0}^\infty \frac{(-1)^b \Gamma(y_{k+2b}^{(m)} - b - 1/2)}{k! b!} e^{\mu_{k+2b}^{(m)} t}
     \frac{y_{k+2b}^{(m)} - 1/2}{2^{2b} \sqrt{\pi}} \\
     \times \Big\{ \sum_{a = 0}^\infty \frac{(k+2(a+b))! \Gamma(a+1/2)}{(2a)! \Gamma(y_{k+2b}^{(m)} + a + 1/2)} \sqrt{ \frac{(k+m+2(a+b))!}{(k+2(a+b))!}} \dbraket{\phi_{k+2(a+b)}^{(m)}}{\rho(t)} \Big\},
\end{gather*}
with the Gamma function $\Gamma(z)$, $y_k^{(m)} = k + m/2$, and $\mu_k^{(m)} = - \kappa_2 [k(k+m+1) + m(m-1)/2]$.
Inserting $\Gamma(a+1/2) = \sqrt{\pi} (2a)!/ (2^{2a} a!)$, $\Gamma(y_{k+2b}^{(m)} - b - 1/2)/\Gamma(y_{k+2b}^{(m)} + a + 1/2) = 1/(y_{k+2b}^{(m)} - b - 1/2)_{a+b+1}$, and $(2(a+b))!/(2^{a+b} (a+b)!) = (2(a+b)-1)!!$ yields
\begin{gather*}
    \dbraket{\phi_k^{(m)}}{\rho(t)} = \sum_{a,b = 0}^\infty (-1)^b \binom{a+b}{b} \sqrt{ \binom{k+m+2(a+b)}{k+m} \binom{k+2(a+b)}{k} } \\
    \times \frac{(2(a+b)-1)!!}{2^{a+b}}  e^{\mu_{k+2b}^{(m)} t}  \frac{y_{k+2b}^{(m)} - 1/2}{(y_{k+2b}^{(m)} - b - 1/2)_{a+b+1}} \dbraket{\phi_{k+2(a+b)}^{(m)}}{\rho(t)}.
\end{gather*}
Introducing $r = a + b$ and $j = b$, we obtain
\begin{equation}
    \begin{gathered}
        \dbraket{\phi_k^{(m)}}{\rho(t)}
        = \sum_{r = 0}^\infty \sqrt{ \binom{k+m+2r}{m+k}\binom{k+2r}{k}} g_{r,k}^{(m)}(t) \dbraket{\phi_{k+2r}^{(m)}}{\rho(0)},
    \end{gathered}
    \label{eq:time.evolv._exact.evolv.}
\end{equation}
with 
\begin{gather*}
    g_{r,k}^{(m)}(t) = \frac{(2r-1)!!}{2^r} \sum_{j = 0}^r (-1)^j \binom{r}{j} e^{\mu_{k+2j}^{(m)} t} \frac{y_{k+2j}^{(m)} - 1/2}{(y_{k+2j}^{(m)} - j - 1/2)_{r+1}}.
\end{gather*}
In the following, we derive \eref{eq:time.evolv._exact.evolv.} using our results.
We first consider more general Liouvillian given by \eref{eq:diag_L}.
We then consider the case with $\omega = U = \kappa_1 = 0$ to see the consistency. 

From the formal solution $\rho(t) = e^{\mathcal{L} t} \rho(0)$, we find 
\begin{equation*}
    \dbraket{\phi_k^{(m)}}{\rho(t)} = \sum_{\mu = - \infty}^\infty \sum_{q = 0}^\infty \dbra{\phi_k^{(m)}} e^{\hat{\mathcal{L}} t} \dket{\phi_q^{(\mu)}} \dbraket{\phi_q^{(\mu)}}{\rho(0)}.
\end{equation*}
The matrix elements of the propagator can be represented using the spectral decomposition \eref{eq:exact_spec.decomp.} as
\begin{equation}
    \begin{gathered}
        \dbra{\phi_k^{(m)}} e^{\hat{\mathcal{L}} t} \dket{\phi_q^{(\mu)}}
        = \sum_{\nu = - \infty}^\infty \sum_{l = 0}^\infty e^{\lambda_l^{(\nu)} t} \dbraket{\phi_k^{(m)}}{\rho_l^{(\nu)}} \dbraket{\bar{\rho}_l^{(\nu)}}{\phi_q^{(\mu)}}.
    \end{gathered}
    \label{eq:time.evolv._prop.me.0}
\end{equation}
Now we need to evaluate $\dbraket{\phi_k^{(m)}}{\rho_l^{(\nu)}}$ and $\dbraket{\bar{\rho}_l^{(\nu)}}{\phi_q^{(\mu)}}$.
To this end, we examine \erefs{eq:diag_Reigvec} and (\ref{eq:diag_Leigvec}).
From the identity ${}_1 F_1(\alpha;\beta; \tau z) e^{z} = \sum_{j = 0}^\infty {}_2 F_1(- j, \alpha; \beta; - \tau) z^n/n!$ with ${}_2 F_1(\alpha,\beta;\gamma;z) = \sum_{j=0}^\infty (\alpha)_j (\beta)_j z^j / ((\gamma)_j j!) \ (\alpha, \beta, \gamma \in \mathbb{C}, \gamma \notin \mathbb{Z}_{\leq 0})$ the hypergeometric function, we find 
\begin{gather*}
    \dket{\rho_l^{(\nu)}} = \sum_{j = 0}^\infty {}_2 F_1 \Big( -j, 1 - x_l^{(\nu)}; 2 - 2 x_l^{(\nu)} - \frac{\kappa_1}{\kappa_2}; 2 \Big) \frac{(- \hat{\mathcal{A}})^j}{j!} \dket{\phi_l^{(\nu)}},
\end{gather*}
and
\begin{gather*}
    \dbra{\bar{\rho}_l^{(\nu)}} = \dbra{\phi_l^{(\nu)}} \sum_{j = 0}^\infty \frac{\hat{\mathcal{A}}^j}{j!}  {}_2 F_1 \Big( -j, x_l^{(\nu)}; 2 x_l^{(\nu)} + \frac{\kappa_1}{\kappa_2}; 2 \Big).
\end{gather*}
These expressions yield
\begin{gather*}
    \dbraket{\phi_k^{(m)}}{\rho_l^{(\nu)}} = \delta_{\nu,m} \ \delta_{l - k, r} \ \delta_{r \geq 0} \sqrt{ \binom{|\nu|+l}{|\nu|+k}\binom{l}{k}} \\
    \times (-1)^r {}_2 F_1 \Big( -r, 1 - x_l^{(\nu)}; 2 - 2 x_l^{(\nu)} - \frac{\kappa_1}{\kappa_2}; 2 \Big),
\end{gather*}
and 
\begin{gather*}
    \dbraket{\bar{\rho}_l^{(\nu)}}{\phi_q^{(\mu)}} = \delta_{\nu,\mu} \ \delta_{q - l, r} \ \delta_{r \geq 0} \sqrt{ \binom{|\nu|+q}{|\nu|+l}\binom{q}{l}} \\
    \times {}_2 F_1 \Big( -r, x_l^{(\nu)}; 2 x_l^{(\nu)} + \frac{\kappa_1}{\kappa_2}; 2 \Big),
\end{gather*}
where $\delta_{r \geq 0} = 0 \ (r < 0)$ and $= 1 \ (r \geq 0)$.
Inserting these into \eref{eq:time.evolv._prop.me.0}, we obtain 
\begin{equation}
    \dbra{\phi_k^{(m)}} e^{\hat{\mathcal{L}} t} \dket{\phi_q^{(\mu)}}
    = \delta_{m,\mu} \ \delta_{q-k,r} \ \delta_{r \geq 0} \sqrt{ \binom{|m|+q}{|m|+k}\binom{q}{k}} \ G_{r,k}^{(m)}(t),
    \label{eq:time.evolv._prop.me}
\end{equation}
with
\begin{equation}
    \begin{gathered}
        G_{r,k}^{(m)}(t) = \sum_{j = 0}^r (-1)^j \binom{r}{j} e^{\lambda_{k+j}^{(m)} t} {}_2 F_1 \Big( -j, 1 - x_{k+j}^{(m)}; 2 - 2 x_{k+j}^{(m)} - \frac{\kappa_1}{\kappa_2}; 2 \Big) \\
        \times {}_2 F_1 \Big( -(r-j), x_{k+j}^{(m)}; 2 x_{k+j}^{(m)} + \frac{\kappa_1}{\kappa_2}; 2 \Big).
    \end{gathered}
    \label{eq:time.evolv._G.def}
\end{equation}
This then yields
\begin{gather*}
    \dbraket{\phi_k^{(m)}}{\rho(t)} = \sum_{r = 0}^\infty \sqrt{ \binom{|m|+k+r}{|m|+k}\binom{k+r}{k}} G_{r,k}^{(m)}(t) \dbraket{\phi_{k+r}^{(m)}}{\rho(0)}.
\end{gather*}

Now we set $\kappa_1 = 0$. In this case, the hypergeometric functions in \eref{eq:time.evolv._G.def} take the form of ${}_2 F_1(-n, \beta; 2\beta; 2)$ with $n$ being a nonnegative integer, which vanishes unless $n$ is even as
\begin{equation*}
    {}_2 F_1(-n, \beta; 2\beta; 2) = \left\{
    \begin{array}{ll}
    \frac{n!}{2^n (n/2)!} \frac{1}{(\beta + 1/2)_{n/2}} & (n: {\rm even})\\
    0 & (n: {\rm odd}).
    \end{array}
    \right.
\end{equation*}
Accordingly, $G_{r,k}^{(m)}(t)$ vanishes unless $r$ is even.
Inserting the above relation of the hypergeometric function in the definition of $G_{2r,k}^{(m)}$, we obtain
\begin{gather*}
    G_{2r,k}^{(m)}(t) = \sum_{j = 0}^r \binom{2r}{2j} \frac{(2j)! (2(r-j))!}{2^{2r} j! (r-j)!} e^{\lambda_{k+2j}^{(m)} t} \frac{1}{(3/2-x^{(m)}_{k+2j})_j (1/2 + x^{(m)}_{k+2j})_{r-j}} \\
    = \frac{(2r-1)!!}{2^r} \sum_{j = 0}^r (-1)^j \binom{r}{j} e^{\lambda_{k+2j}^{(m)} t} \frac{x_{k+2j}^{(m)} - 1/2}{(x_{k+2j}^{(m)} - j - 1/2)_{r+1}},
\end{gather*}
where we have used $\binom{2r}{2j} (2j)! (2(r-j))! / (2^{2r} j! (r-j)! ) = \binom{r}{j} (2r-1)!!/2^r$ and $[(3/2-z)_j (1/2+z)_{r-j}]^{-1} = (-1)^j (z-1/2)/(z-j-1/2)_{r+1}$.
When we further set $U = \omega = 0$, it follows that $x_k^{(m)} = k + m/2 = y_k^{(m)}$ and $\lambda_k^{(m)} = - \kappa_2 [k(k+m+1) + m(m-1)/2] = \mu_k^{(m)}$. Therefore, it follows that $G_{2r,k}^{(m)}(t) = g_{r,k}^{(m)}(t)$ and \eref{eq:time.evolv._exact.evolv.} is reproduced using our results.

Using the above results, we can also express the Heinsenberg representation of an operator $O$ defined by $\dket{O^{H} (t)} = e^{\hat{\mathcal{L}}^\dagger t} \dket{O}$.
In the Fock basis, we find 
\begin{gather*}
    \dbraket{\phi_k^{(m)}}{O^H(t)} = \sum_{\mu = - \infty}^\infty \sum_{q = 0}^\infty \dbra{\phi_k^{(m)}} e^{\hat{\mathcal{L}}^\dagger t} \dket{\phi_q^{(\mu)}} \dbraket{\phi_q^{(\mu)}}{O}.
\end{gather*}
From \eref{eq:time.evolv._prop.me}, it follows that
\begin{equation*}
    \begin{gathered}
        \dbra{\phi_k^{(m)}} e^{\hat{\mathcal{L}}^\dagger t} \dket{\phi_q^{(\mu)}} = \dbra{\phi_q^{(\mu)}} e^{\hat{\mathcal{L}} t} \dket{\phi_k^{(m)}}^*
        = \delta_{m,\mu} \ \delta_{k-q \geq 0} \ \sqrt{ \binom{|m|+k}{|m|+q}\binom{k}{q}} \ G_{k-q,q}^{(-m)}(t),
    \end{gathered}
\end{equation*}
where we have used $[G_{r,k}^{(m)}(t)]^* = G_{r,k}^{(-m)}(t)$.
Inserting this, we obtain 
\begin{gather*}
    \dbraket{\phi_k^{(m)}}{O^H(t)} = \sum_{q = 0}^k \sqrt{ \binom{|m|+k}{|m|+q}\binom{k}{q}} G_{k-q,q}^{(-m)}(t) \dbraket{\phi_q^{(m)}}{O}.
\end{gather*}

\end{document}